\documentclass[journal=jpccck,manuscript=article,layout=twocolumn]{achemso}

\usepackage{float}
\usepackage{graphicx}
\usepackage{epstopdf}
\usepackage{booktabs}
\usepackage{chemformula}
\usepackage[T1]{fontenc}
\usepackage[utf8]{inputenc}
\usepackage{siunitx}
\DeclareSIUnit\angstrom{\text{Å}}
\usepackage[super]{nth}
\usepackage{dsfont}
\SectionNumbersOn

\author{Sina Safaei}
\email{sina.safaei@ugent.be}
\affiliation{BioMMedA, Institute of Biomedical Engineering, Ghent University, 9000 Gent, Belgium}

\author{Parham Rezaee}
\affiliation{BioMMedA, Institute of Biomedical Engineering, Ghent University, 9000 Gent, Belgium}

\author{An Ghysels}
\email{an.ghysels@ugent.be}
\affiliation{BioMMedA, Institute of Biomedical Engineering, Ghent University, 9000 Gent, Belgium}

\title[HRETIS]{Collaborate to decorrelate in path space: Hamiltonian replica exchange transition interface sampling (HRETIS)}


\begin{document}
\twocolumn[
\begin{@twocolumnfalse}
\begin{abstract}
We present Hamiltonian Replica Exchange Transition Interface Sampling (HRETIS), a path sampling framework designed to efficiently sample rare events in systems with complex potential energy landscapes. HRETIS introduces a helper potential within a Hamiltonian replica exchange scheme, which enhances exploration of path space when the underlying potential is not well suited for conventional path sampling approaches. This is particularly advantageous for systems exhibiting multiple pathways separated by orthogonal barriers such as in drug (un)binding, where standard algorithms often show slow convergence since they become trapped within specific pathways. 
By exchanging Hamiltonians between the path ensembles, HRETIS overcomes these limitations and increases the decorrelation between subsequent paths in the Monte Carlo chain.
We demonstrate that HRETIS provides robust and accurate kinetics in several systems, including coarse-grained simulations of amino acid permeation through a dipalmitoylphosphatidylcholine (DPPC) membrane. 
Moreover, HRETIS is found to improve sampling efficiency and convergence,
illustrating its potential as a powerful tool for rare event sampling in complex molecular systems.
\end{abstract}
\end{@twocolumnfalse}
]

\section{Introduction}

Molecular dynamics (MD) simulations of biomolecular systems rarely reach timescales beyond hundreds of microseconds, far below the millisecond to second regime required for rare events such as drug (un)binding,\cite{plattner2015protein} protein-protein association kinetics,\cite{plattner2017complete} membrane transport,\cite{frallicciardi2022determining} lipid translocation,\cite{martinotti2020molecular, elber2020milestoning} and protein conformational changes \cite{wang2022efficient, li2025enhanced, haloi2025adaptive}. This timescale gap prevents reliable estimation of kinetics and has motivated the development of path sampling methodologies that capture rare events. Over the past two decades, these methods have evolved significantly to estimate rates and committors,\cite{wang2023predicting, kang2024computing} and are nowadays often assisted by machine learning techniques \cite{tsai2022path, asghar2024efficient, casert2024learning, post2025ai, trizio2025everything, horvath2025stim1}. The original formulation, transition path sampling (TPS), is a Monte Carlo (MC) approach designed to sample ensembles of trajectories between stable states without modifying the underlying dynamics. \cite{dellago1998transition} In TPS, new trajectories are generated through the \emph{shooting} move, where the velocity of a phase point in an existing trajectory is 
modified and then propagated in time using MD to produce a new trial trajectory. \cite{dellago1998efficient} The trial trajectory then is accepted or rejected according to a Metropolis-Hastings acceptance criterion that enforces detailed balance in path space.\cite{bolhuis2002transition, dellago2008transition}

\begin{figure}[b!]
    \includegraphics[width=\columnwidth]{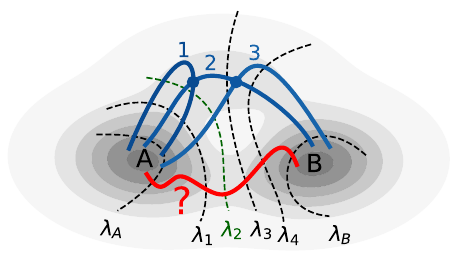}
    \caption{
    Schematic illustration of TIS in ensemble $[2^+]$ for a simplified two-channel system.
    Subsequent trajectories in a path sampling Monte Carlo (MC) chain can be too similar.
    E.g.\ the three blue trajectories are strongly correlated and all sample the upper channel. More path-space exploration is needed to sample trajectories in the lower channel.}
    \label{fig:intro}
    \end{figure}

Transition interface sampling (TIS) computes rates and crossing probabilities through a series of path sampling simulations.\cite{van2003novel, van2007reaction} TIS introduces an order parameter $\lambda$ describing the transition between two stable states A and B, along which a set of $n+1$ interfaces $\lambda=\lambda_i$ ($i=0,..,n$) is constructed. 
Trajectories are then sampled conditioned on making progress along $\lambda$, where each simulation targets a distinct path ensemble $[i^+]$, corresponding to a different stage of the transition. Path ensemble $[i^+]$ is defined as the ensemble of trajectories that start at $\lambda_A$, cross interface $\lambda_i$, and terminate either upon reaching $\lambda_B$ or returning to $\lambda_A$ (see Fig.~\ref{fig:intro}).

Correlations between consecutively sampled trajectories remain a common challenge in path sampling, which may lead to slow convergence of the simulations. \cite{vervust2025path, vervust2026estimating} This is particularly challenging in systems with multiple reaction channels. An example is the rare event permeation through a heterogeneous phospholipid membrane, e.g.\ with raft domains, where the mechanism and kinetics will depend on which pathway the drug molecules takes to permeate. Decorrelation of paths in this context refers to the ability of the MC chain to sample multiple permeation pathways, as illustrated in Fig.~\ref{fig:intro}.

The correlation challenge has been discussed in literature \cite{bolhuis2015practical, chong2017path, hall2022practical, van2023far} and addressed through several algorithms specifically designed to enhance exploration of path space and improve decorrelation. One approach is replica exchange transition interface sampling (RETIS) introduced by van Erp et al., where paths are exchanged between adjacent path ensembles $[i^+]$ and $[(i+1)^+]$ with the \textit{replica exchange} move \cite{van2007reaction}. These exchange moves are computationally cost-free as they do not require generation of new trajectories, while still enhancing decorrelation between consecutively sampled paths in the MC chain. This effect was later strengthened by the \textit{zero-swap} move, which introduces a $[0^-]$ ensemble that explores only the reactant state A. Paths are then exchanged between the $[0^-]$ ensemble ($\lambda<\lambda_A$) and the $[0^+]$ ensemble ($\lambda>\lambda_A$), which improves sampling of the reactant basin before re-entry into the barrier region \cite{bolhuis2008rare}.
The replica exchange move was further extended to the limit of infinite swapping between all possible ensembles in the \textit{infinite-swap} move of the $\infty$RETIS framework, which leads to improved decorrelation \cite{roet2022exchanging, zhang2024highly}.
Recently, Fujisaki et al.\ put forward exchange of paths between ensembles at different temperatures within a framework restricted to overdamped dynamics \cite{fujisaki2010onsager}.
In configurational space, exchange-based strategies have also been developed to improve exploration efficiency in complex energy landscapes. These include the replica-exchange method
between replicas at different temperature,\cite{hukushima1996exchange, sugita1999replica}
different Hamiltonians or thermodynamic potentials,\cite{sugita2000multidimensional, okabe2001replica, bussi2014hamiltonian, Chen2016multiple}
and between parallel nested simulations \cite{unglert2025replica}.

Another class of approaches involves improving shooting algorithms by introducing novel shooting-based moves such as the \textit{web-throwing},\cite{riccardi2017fast} \textit{stone-skipping},\cite{riccardi2017fast} and \textit{wire-fencing}\cite{zhang2023enhanced} move. These moves reduce correlations between successive trajectories in the MC chain by generating intermediate subtrajectories. 
Another improvement was in the selection of shooting points, for example using conditioned Boltzmann generators to draw shooting points directly from the relevant region of configuration space, which eliminates dependence on previously generated trajectories \cite{noe2019boltzmann, falkner2023conditioning}. A related data-driven approach is reactivity-biased shooting, where shooting points are selected in a low-dimensional collective variable space learned from trajectory data. \cite{zhang2024combining}
Alternative methods combine metadynamics with shooting moves in TPS, where a history-dependent bias constructed from Gaussian potentials deposited on previously sampled trajectories, is used during trajectory generation to promote transitions between reaction channels while preserving unbiased dynamics along the generated paths \cite{borrero2016avoiding, bolhuis2021transition}. Further developments explore a methodology in which configuration space and path space are sampled in parallel using metadynamics and TPS, with exchanges between configuration and path ensembles.\cite{falkner2024enhanced} This enhances barrier crossing and reduces correlations through relaxation in configuration space, while leaving the dynamics along accepted trajectories unbiased.

All of the above mentioned methods attempt to enhance the decorrelation of the sampled trajectories to aid the sampling efficiency in path space.
Here, we introduce Hamiltonian replica exchange transition interface sampling (HRETIS), where exchange is made possible between paths simulated under different Hamiltonians.
The core idea of HRETIS is that trajectories are generated independently under each Hamiltonian, and at a chosen exchange frequency, an \textit{engine-swap} move is performed: phase points are randomly selected from each path and exchanged between the Hamiltonians (the MD engines) using a Metropolis–Hastings acceptance criterion.
The Hamiltonians could differ, for example, in the degree of hydrophobicity of the molecule or another specific region of interest. 
HRETIS is expected to be most useful when the main Hamiltonian $\mathcal{H}_\text{m}$ is higher-cost or slower-exploring, whilst the helper Hamiltonian $\mathcal{H}_\text{h}$ is lower-cost and faster-exploring.
In such a setup, the helper Hamiltonian accelerates path decorrelation for the main Hamiltonian, i.e.\ enhancing exploration of path space, and improves convergence. 
The use of different Hamiltonians in path sampling is conceptually related to the QuanTIS approach, in which the $[0^-]$ paths are propagated at a lower level of theory (e.g.\ classical force field) than the $[i^+]$ paths (e.g.\ quantum mechanics)
to reduce the overall computational cost.\cite{lervik2015gluing}
While QuanTIS uses only one Hamiltonian in each ensemble, HRETIS allows both a main and helper Hamiltonian within a single ensemble.

This paper presents the HRETIS algorithm and provides its detailed balance derivation. The method is first validated on simple one-dimensional model systems. Improvements in path-space sampling are then examined using two-dimensional test cases with two channel permeation pathways. Finally, path decorrelation and convergence behavior are assessed by studying drug permeation across a phospholipid membrane at the coarse-grained scale.

\section{Results \label{Sec:results}}

\subsection{HRETIS algorithm}
\label{sec:HRETIS_algorithm}

\begin{figure*}[htb]
    \includegraphics[width=\linewidth]{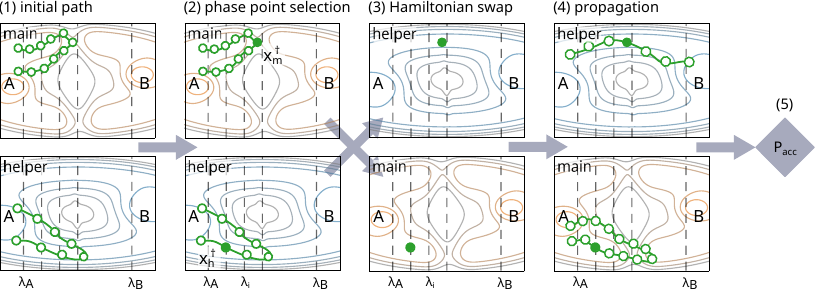}
    \caption{The HRETIS algorithm. (1) The procedure begins with one old trajectory in each system, denoted by $X_{\text{m}}^{\text{(o)}}$ and $X_{\text{h}}^{\text{(o)}}$, corresponding to the main and helper Hamiltonians, respectively. 
    (2) A phase point is selected randomly from each trajectory, denoted by
    $x_\text{m}^\dagger$ and $x_\text{h}^\dagger$.
    (3) The Hamiltonians are swapped.
    (4) Backward and forward integrations are performed from the selected phase points to generate new trajectories, $X_{\text{m}}^{\text{(n)}}$ and $X_{\text{h}}^{\text{(n)}}$.
    (5) The generated paths are accepted or rejected according to the acceptance criterion in Eq.~\ref{eq:acc-criterion-method}.
    }
	\label{Fig:schematic}
\end{figure*}

HRETIS extends the RETIS framework to multiple Hamiltonians. The core of the algorithm is shown schematically in Fig.~\ref{Fig:schematic}. Assume the subscripts `m' and `h' refer to the main and helper Hamiltonians $\mathcal{H}_\text{m}$ and $\mathcal{H}_\text{h}$ with potential energies $U_\text{m}$ and $U_\text{h}$. The symbol $X$ denotes a path, while $x$ represents a phase point consisting of the positions and momenta.
The superscripts (o) and (n) indicate the old and new paths $X^\text{(o)}$ and $X^\text{(n)}$ in the MC move, respectively. 
Concepts of path sampling with RETIS are reviewed shortly in the Methods Section~\ref{sec:retis-review}.

A superstate $z$ is introduced to describe the overall state of the main and helper Hamiltonians simultaneously, $z=(X_\text{m},X_\text{h})$. 
Further, $\beta = (k_B T)^{-1}$, and the indicator function $\mathds{1}_{i}[X]$ is defined as 1 if path $X$ belongs to ensemble $[i^+]$ and 0 otherwise. 

The Hamiltonian exchange move begins with selecting an ensemble $[i^+]$ randomly. The exchange is also referred to as an engine-swap move, since in practice it concerns the MD integration `engines'.
\begin{enumerate}
\item The move starts from the old state $z^\text{(o)}$ with old paths
$X_\text{m}^{\text{(o)}}$ and $X_\text{h}^{\text{(o)}}$ that are the most recently accepted paths for each Hamiltonian in $[i^+]$, with path lengths $N_\text{m}^{\text{(o)}}$ and $N_\text{h}^{\text{(o)}}$ (expressed in number of phase points), respectively. 

\item Phase points $x_\text{m}^\dagger$ and $x_\text{h}^\dagger$ are selected randomly from the paths $X_\text{m}^{\text{(o)}}$ and $X_\text{h}^{\text{(o)}}$, respectively. Their potential energies under the main or helper Hamiltonian are $U_\text{m}(x_\text{m}^\dagger)$ and $U_\text{h}(x_\text{h}^\dagger)$.

\item The engines are swapped. Phase points $x_\text{m}^\dagger$ and $x_\text{h}^\dagger$ are now simulated under swapped Hamiltonians, i.e.\ under the helper and main Hamiltonians, respectively. Their potential energies under the swapped Hamiltonians are $U_\text{h}(x_\text{m}^\dagger)$ and $U_\text{m}(x_\text{h}^\dagger)$.

\item The selected phase points $x_\text{m}^\dagger$ and $x_\text{h}^\dagger$ are propagated both backward and forward in time under the swapped Hamiltonians, to generate new paths $X_\text{h}^{\text{(n)}}$ and $X_\text{m}^{\text{(n)}}$, with path lengths $N_\text{h}^{\text{(n)}}$ and $N_\text{m}^{\text{(n)}}$.

\item The new state $z^\text{(n)}= (X_\text{m}^{\text{(n)}}, X_\text{h}^{\text{(n)}})$ with the generated paths is accepted or rejected according to the Metropolis acceptance criterion (derivation in Appendix \ref{sec:DetailedBalance}),
\begin{multline}
P_{\text{acc}} =
\mathds{1}_{i}[X_{\text{m}}^{\text{(n)}}]
\mathds{1}_{i}[X_{\text{h}}^{\text{(n)}}] \\
\times \min \left\{ 
1, \frac{N_{\text{m}}^{\text{(o)}} N_{\text{h}}^{\text{(o)}}}{N_\text{m}^{\text{(n)}} N_\text{h}^{\text{(n)}}}
\; e^{-\beta\Delta\Delta U } \right\}
\label{eq:acc-criterion-method}
\end{multline}
\end{enumerate}
The term $\Delta\Delta U$ is the difference between helper and main Hamiltonians in energy shifts when the phase points are swapped,
\begin{multline}
\Delta\Delta U = U_{\text{h}}(x_{\text{m}}^\dagger) - U_{\text{h}}(x_{\text{h}}^\dagger) \\
- U_{\text{m}}(x_{\text{m}}^\dagger) + U_{\text{m}}(x_{\text{h}}^\dagger)
\label{Eq:DeltaDeltaU}
\end{multline}
Rearranging the terms in $\Delta\Delta U$ shows that swapping phase points is equivalent to swapping Hamiltonians. This term is identical to the expression obtained for a zero-swap move between two levels of theory (e.g.\ quantum mechanical versus classical), as implemented in QuanTIS. \cite{lervik2015gluing}

The acceptance criterion is based on three aspects: whether each path remains a member of the $[i^+]$ ensemble, the path lengths, and $\Delta\Delta U$.
If the new paths are rejected in step 5, the engine-swap is a fairly expensive MC move, of the same order as a standard shooting move in both ensembles because of the MD propagation in step 4. For this reason, instead of evaluating the acceptance criterion after completing the full Hamiltonian exchange move, it is decomposed into sequential checks. Indeed, the following modified acceptance criterion also ensures detailed balance,
\begin{multline}
P'_{\text{acc}} =
\mathds{1}_{i}[X_{\text{m}}^{\text{(n)}}]
\mathds{1}_{i}[X_{\text{h}}^{\text{(n)}}] \\
\times 
\min \left\{ 
1, \frac{N_{\text{h}}^{\text{(o)}}}{N_\text{h}^{\text{(n)}}} \right\}
\min \left\{ 
1, \frac{N_{\text{m}}^{\text{(o)}}}{N_\text{m}^{\text{(n)}}} \right\}
\; P^{\Delta\Delta U}_{\text{acc}} 
\label{eq:acc-criterion-method2}
\end{multline}
with
\begin{equation}
P^{\Delta\Delta U}_{\text{acc}} = \min \left\{ 1, e^{-\beta \Delta\Delta U } \right\}
\label{eq:acc-criterion-DDU}
\end{equation}
The modified acceptance criterion $P'_{\text{acc}}$ contains five factors that are evaluated throughout the implementation of HRETIS. Rejection by any factor leads to a rejection of the Hamiltonian exchange move and recounting of the old paths. The next action in the algorithm is only started upon acceptance of the previous factor. When all factors are accepted, the old paths are effectively replaced with the new paths.

The most important improvement in the computational efficiency of the HRETIS algorithm is the acceptance criterion $P^{\Delta\Delta U}_{\text{acc}}$ defined in Eq.~\ref{eq:acc-criterion-DDU}. Evaluating $\Delta\Delta U$ of Eq.~\ref{Eq:DeltaDeltaU} requires the potential energies under both Hamiltonians for the two selected phase points $x_\text{m}^\dagger$ and $x_\text{h}^\dagger$. Compared to MD integration, this evaluation is computationally very cheap, especially for classical force fields. The early rejection between steps 3 and 4 of the algorithm (Fig.~\ref{Fig:schematic}) thus avoids unnecessary MD integration for paths that would likely be rejected later. This decomposition is one of the key structural advantages of the HRETIS algorithm.

The HRETIS algorithm is implemented in the \texttt{infRETIS} code\cite{infretissoftware} (flowchart in Supplementary Fig.~S1). Additional improvements that increase computational efficiency are discussed in Section~\ref{sec:improvements}, e.g.\ concerning the random phase point selection and the wire-fencing move.

\subsection{Demonstration of HRETIS with 1D potentials}

\begin{figure}[tbh!]
    \includegraphics[width=\columnwidth]{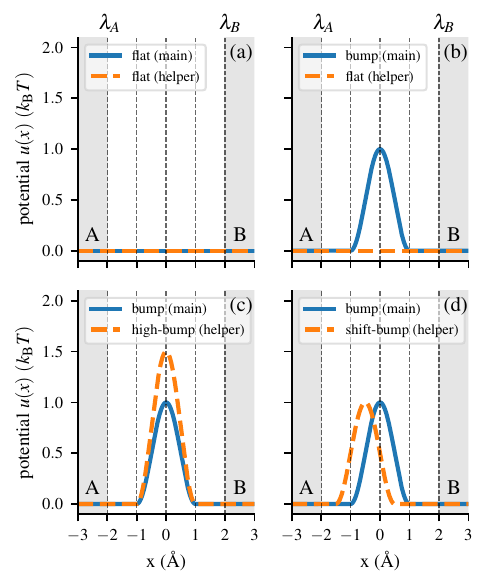}
    \includegraphics[width=\columnwidth]{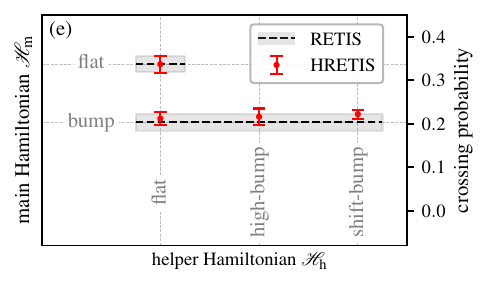}
    \caption{Demonstration of the HRETIS methodology with 1D potentials $u(x)$. (a) Simplest validation case where both main and helper Hamiltonians are flat potentials. For (b)–(d), the main potential $\mathcal{H}_\text{m}$ is a cosine-bump barrier (bump), while the helper Hamiltonian $\mathcal{H}_\text{h}$ is (b) flat, (c) a cosine-bump barrier with increased height (high-bump), and (d) a shifted cosine-bump barrier (shift-bump). Gray areas indicate states A and B; interfaces $\lambda_i$ indicated by vertical dashed lines.
    (e) Crossing probability $P_A(\lambda_B|\lambda_A)$ of $\mathcal{H}_\text{m}$. 
    Error bars for HRETIS and shaded gray region for RETIS are two standard error estimates.
	\label{Fig:1d_potentials}}
\end{figure}

To demonstrate the newly implemented HRETIS methodology, an argon-like particle was simulated with Langevin dynamics on a simple one-dimensional potential $u(x)$. The main Hamiltonian $\mathcal{H}_\text{m}$ had either a flat potential or a cosine-shaped (`bump') energy barrier (Fig.~\ref{Fig:1d_potentials}a-d, blue), similarly to earlier work \cite{ghysels2021exact,vervust2026estimating}. While a flat potential does not cause a rare event, it is a straight-forward validation test system. RETIS without engine-swap was therefore used to create a benchmark of the crossing probability $P_A(\lambda_B|\lambda_A)$ for $\mathcal{H}_\text{m}$, corresponding to transitions from the left side of the potential to the right side (Fig.~\ref{Fig:1d_potentials}e).

Next, HRETIS was employed with different helper Hamiltonians $\mathcal{H}_\text{h}$: a flat potential, a cosine-bump barrier with increased height (`high-bump'), and a cosine-bump barrier shifted relative to the main barrier (`shift-bump'). The potentials $u(x)$, along with the interfaces $\lambda_i$, are shown in Fig.~\ref{Fig:1d_potentials}a-d. HRETIS was performed using an engine-swap probability of 50\%. 

In three cases (Fig.~\ref{Fig:1d_potentials}b-d), the helper Hamiltonian $\mathcal{H}_\text{h}$ has a non-perfect overlap in its Boltzmann distribution with the main Hamiltonian $\mathcal{H}_\text{h}$. Nevertheless, as shown in Fig.~\ref{Fig:1d_potentials}e, the HRETIS crossing probability $P_A(\lambda_B|\lambda_A)$ belonging to the main Hamiltonian is in excellent agreement with the RETIS benchmark for each studied main-helper combination. 

Regarding computational efficiency, the acceptance ratio in HRETIS is lower than in RETIS due to the engine-swap contribution of Eq.~\ref{eq:acc-criterion-method}. In RETIS, the flat and bump potentials yield acceptance ratios of 54\% and 49\%, respectively, whereas in HRETIS these decrease to 45\% and 39\%. This implies that more MC cycles are required in HRETIS to generate the same number of accepted paths $N_\text{acc}$. However, when the rejection is caused by the $\Delta\Delta U$ term in Eq.~\ref{eq:acc-criterion-DDU}, only two phase points need to be evaluated, while the average path length could be on the order of thousands of phase points. This makes the computational cost of rejections due to $\Delta\Delta U$ in Eq.~\ref{eq:acc-criterion-DDU} negligible.

\subsection{Enhanced path decorrelation in a 2D potential
\label{sec:results2d}}

\subsubsection{Model membrane with 2 permeation channel}

\begin{figure*}[htb!]
    \includegraphics[width=1.0\linewidth]{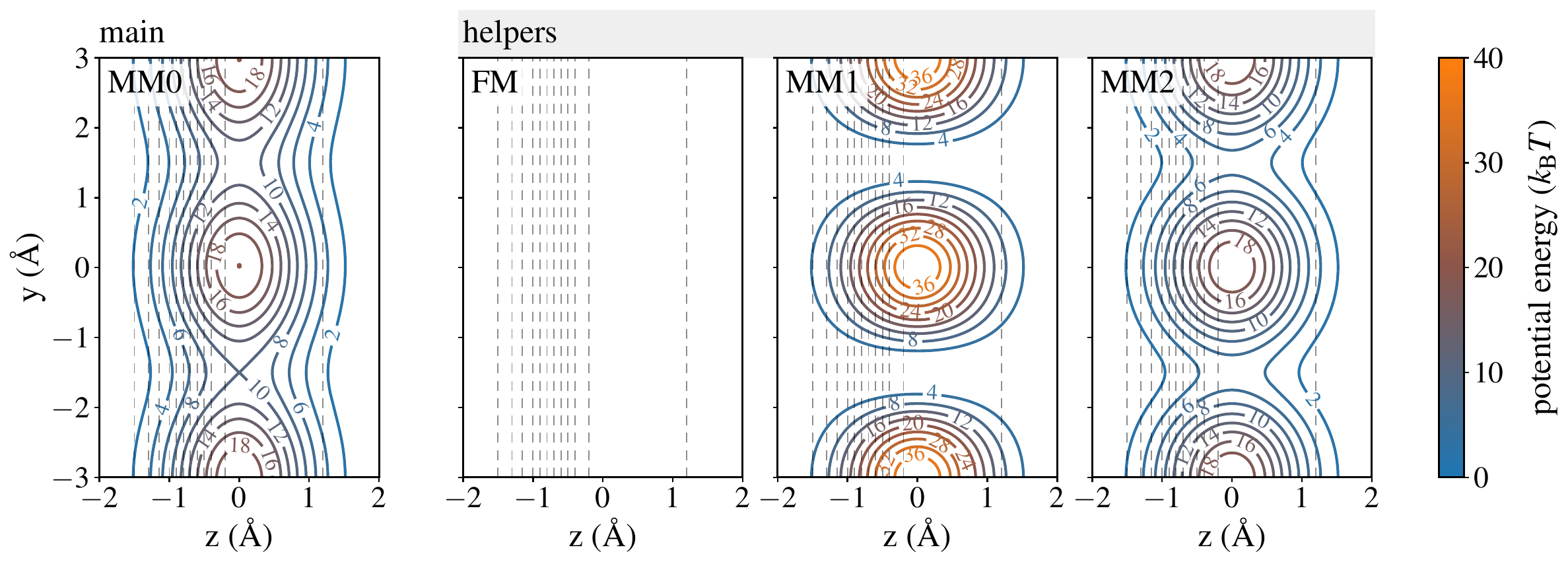}
    \caption{The potential $V(y,z)$ in the main Hamiltonian $\mathcal{H}_\text{m}$ represents a membrane with two permeation channels (MM0). The main Hamiltonian is combined with a helper potential in $\mathcal{H}_\text{h}$, either a flat membrane (FM), membrane model 1 (MM1), or membrane model 2 (MM2).
    The contour lines for MM1 have sparser increments for readability.
    Interfaces indicated by vertical dashed lines.}
	\label{Fig:2d_potentials}
\end{figure*}

To investigate whether HRETIS enhances phase space exploration, we employed a two-dimensional potential energy surface $V(y,z)$ (Fig.~\ref{Fig:2d_potentials}) that represents a simplified model of an argon-like particle permeating through a lipid bilayer with two separate permeation channels, as introduced in earlier work\cite{ghysels2021exact}. The barrier in the upper and lower channel of this membrane model (MM0) are 11\,$k_BT$ and 10\,$k_BT$, respectivly.
It is a typical example of the decorrelation challenge of Fig.~\ref{fig:intro} where the kinetics from A to B can occur through two distinct pathways, here an upper and lower channel.
The MM0 potential was used for the main Hamiltonian $\mathcal{H}_\text{m}$, while three additional potentials (Fig.~\ref{Fig:2d_potentials}) were proposed for the helper Hamiltonians $\mathcal{H}_\text{h}$: (1) a flat membrane (FM) that allows free permeant motion, (2) a model membrane 1 (MM1) whose permeation pathways have no energy barrier, 
and (3) a model membrane 2 (MM2) whose pathways have only a small energy barrier. 

The MC chain in RETIS (and HRETIS) is somewhat complex. 
In the present RETIS simulation with 12 ensembles, 12 MC chains are sampled simultaneously. Within a chain, each new accepted path is generated through a standard shooting move. This is immediately followed by the infinite-swap move of $\infty$RETIS which can change the ensemble index $[i^+]$ of this new accepted path.
HRETIS also simulates 12 MC chains simultaneously, but unlike RETIS, it offers two options in every MC cycle to generate a new path under $\mathcal{H}_\text{m}$: the standard shooting move or our new Hamiltonian exchange move. The latter is executed with a certain engine-swap probability (e.g.\ s50 refers to 50\% engine-swap frequency).
Moreover, there are also 12 extra MC chains in HRETIS under $\mathcal{H}_\text{h}$, for which 
we implemented that a shooting move is performed with a certain probability
in order to explore the helper potential (see improvements in Section~\ref{sec:improvements}).

\subsubsection{Channel switching}

An important aspect to obtain accurate kinetics is whether both the upper and lower permeation channels are sampled sufficiently. If an MC chain of paths is stuck in a single channel, the decorrelation of the paths is insufficient (cfr.\ Fig.~\ref{fig:intro}). Therefore a similar analysis is performed as in Ref.~\citenum{ghysels2021exact}, by tracing the first crossing point (FCP) of each path. The FCP of a path in ensemble $[i^+]$ is the first phase point $(y^*, z^*)$ on that path immediately after crossing $\lambda_i$.
The value of $y^{*}$ thus provides an indicator of the permeation channel visited by the path.
If the distribution of $y^*$ covers the whole range $[-3,3]$ in Fig.~\ref{Fig:2d_potentials}, then the MC chain has adequately sampled both channels.

\begin{figure*}[!ht]
    \includegraphics[width=\textwidth]{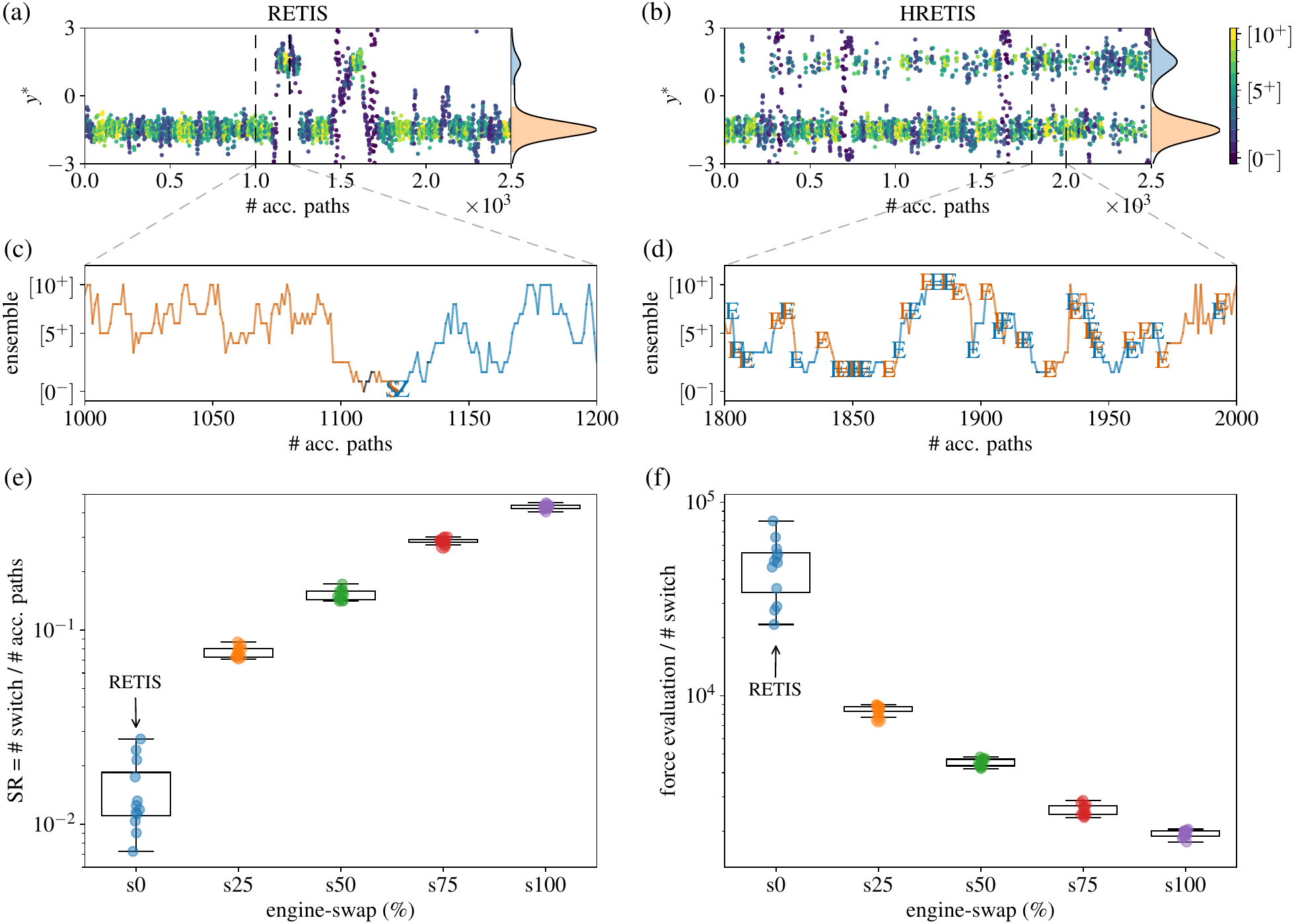}
    \caption{Channel switching in the model membrane (MM0) 2D potential $V(y,z)$.
    HRETIS simulations with MM2 as the helper Hamiltonian. 
    (a) RETIS and (b) HRETIS (s50): $y^*$ of the FCP for each accepted path in MC chain 12, colored by ensemble index, along with the $y^*$ distribution (no MC weights taken into account) on the right (upper channel blue, lower channel orange)
    (c-d) Mechanism of channel switches. Each channel switch is labeled S, Z, and E, corresponding to shooting, zero-swap, and engine-swap moves, respectively. 
    Ensemble index of the accepted path is colored according to whether it is in the upper channel (blue), lower channel (orange), or non-channel regions (when $|y^*|\notin [0.5,2.5]$, black).
    (e) Switching ratio $SR$ and (f) switching cost as a function of engine-swap probability (0\% to 100\%) for all 12 MC chains.}
	\label{Fig:2d_analysis}
\end{figure*}

Figs.~\ref{Fig:2d_analysis}a-b show the channel sampling on the MM0 potential in one arbitrary MC chain (chain 12) of the RETIS and HRETIS (s50, MM2 helper) simulations. 
The $y^{*}$ for the first 2500 accepted (unique) paths in the chain are colored according to the ensemble index $[i^+]$ in which the path was accepted.
Note that the asynchronous infinite-swap of $\infty$RETIS causes the ensemble indices to jump along the chain.
The $y^*$ values show that the RETIS chain starts in the lower channel and requires a large number of new paths before it starts sampling the upper channel. Meanwhile, HRETIS readily explores both channels without becoming trapped in a localized region of phase space.

Looking at the other chains and the simulations with the other helper Hamiltonians FM or MM1 (Supplementary Figs.~S3--S6), a RETIS chain can be trapped in a channel for a high number of MC cycles, 
wheres the Hamiltonian exchange in HRETIS enables channel switches in all ensembles. 
All three helper Hamiltonians (FM, MM1, MM2) in HRETIS could thus increase the decorrelation in the trajectories compared to RETIS.

Figs.~\ref{Fig:2d_analysis}c-d show the mechanisms responsible for switching between channels, with a marker indicating the move responsible for each switch, which appears whenever an accepted path switches from one channel to another. This allows to identify which move typically causes a channel switch and the ensemble in which it occurs. 
In the RETIS simulation, channel switches occur exclusively in the lower ensembles (sampling close to state A) and are triggered by shooting or zero-swap moves. 
The mechanism to switch to the other channel is thus that the MC chain needs to explore the neighborhood of reactant state A (the lower ensembles) where paths can move more freely in the $y$-direction. The enhanced decorrelation in the coordinate $y$, orthogonal to $\lambda$, 
was indeed the original motivation to introduce the $[0^-]$ ensemble covering state A in the RETIS methodology.\cite{van2007reaction}

In contrast, in the HRETIS simulation, paths can switch between channels in all ensembles and predominantly via engine-swap moves. The mechanism of channel switching is thus much more diverse with HRETIS than with RETIS, as it can occur in many ensembles. 
This is the crucial strength of HRETIS: the Hamiltonian exchange has the power to enhance decorrelation in path space across all ensembles.

\subsubsection{Switching ratio}

The efficiency of the channel switches is quantified by the switching rate $SR$, which is defined as the ratio of the number of channel switches to the number of accepted paths $N_\text{acc}$ in an MC chain. The switching rate $SR$ is shown for the 12 MC chains in Fig.~\ref{Fig:2d_analysis}e and Supplementary Fig.~S7. For each helper Hamiltonian, 
increasing the percentage of engine-swap moves
(s0 to s100) leads to an increased $SR$.
This indicates that the Hamiltonian exchange was consistently more effective in generating decorrelated paths for the membrane model potential MM0.

Comparison between different helper Hamiltonians $\mathcal{H}_\text{h}$ further reveals that a higher $SR$ was obtained when the helper potential more closely resembles the main potential. 
This can be understood as follows. 
On one hand, a vastly different helper Hamiltonian can enhance the phase space exploration. For instance, the FM helper promotes broad sampling in the whole $yz$-plane. On the other hand, it can also reduce the acceptance based on the $P_\text{acc}^{\Delta\Delta U}$ criterion of Eq.~\ref{eq:acc-criterion-DDU}, as many more unfavorable $\Delta\Delta U$ values (Eq.~\ref{Eq:DeltaDeltaU}) would be encountered. While the double difference $\Delta\Delta U$ assists in canceling most of the energy differences between the $x^\dagger_\text{m}$ and $x^\dagger_\text{h}$ phase points in the barrier direction along $\lambda$, it cannot fully annihilate energy differences caused by the orthogonal coordinate $y$.

This is a known challenge in alchemical transformations for free energy calculations where sufficient overlap between Boltzmann distribution is needed for feasibility and accuracy.\cite{henin2022enhanced, zhang2024alchemical, robo2023fast} Here, for kinetics calculations rather than free energy calculations, a decent overlap in Boltzmann distributions of the main and helper Hamiltonians is still desirable, while diversity in the Hamiltonians is also needed to sample decorrelated trajectories. Interestingly, a relatively small contribution of accepted engine-swaps can already raise the channel switches efficiency drastically.
For instance, the FM potential is markedly different from the MM0 potential, and only 7\,\% of the accepted cycles for the FM Hamiltonian originated from a successful engine-swap move in the s50 simulation. Yet, this low number of accepted engine-swaps already increased the average $SR$ to $4.7\times10^{-2}$ compared to $1.5\times10^{-2}$ for RETIS.
Generally speaking, for higher dimensional systems, we think it is needed to be cautious with definite statements about how distinctly different helper Hamiltonians can assist path decorrelation, as we suspect it depends on the system-dependent balance between the more adventurous sampling versus the higher rejection rate of the engine-swaps.

Lastly, we implemented an additional move to promote the decorrelation of the paths under the helper Hamiltonian $\mathcal{H}_\text{h}$. The \textit{helper exploration} move 
consists of shooting move or wire-fencing move
in the $\mathcal{H}_\text{h}$ ensembles 
and allows for their sampling independently of the main Hamiltonian.
As seen in Supplementary Fig.~S7, increasing the frequency of helper exploration 
does not significantly affect $SR$. Nevertheless, helper exploration is expected to 
be valuable in applications where the helper Hamiltonian is computationally very cheap, and where it can assist in reaching other regions of the path space before being exchanged back to the main Hamiltonian.

\subsubsection{Computational efficiency}

As a measure of computational efficiency, the number of force evaluations (or MD steps) required to generate a certain number of accepted paths $N_\text{acc}$ can be used.\cite{falkner2024enhanced} 
This accounts for the computational cost of all generated paths, encompassing both accepted and rejected paths
and therefore provides a fair comparison of the computational cost of RETIS and HRETIS.
Here, to highlight the path decorrelation, we measure the cost as the average number of force evaluations required per channel switching event. This cost will be high in case the MC chain has many rejected paths or in case the paths in the chain are highly correlated and trapped in a specific channel without switching.
Fig.~\ref{Fig:2d_analysis}f shows the cost 
for RETIS and HRETIS (s50) with MM2 as the helper Hamiltonian. 
The trend shows that an increase in engine-swap probability leads to a decrease in computational cost.
This trend is also observed for the other helper Hamiltonians FM and MM1 (Supplementary Figs.~S8). This is a convincing result promoting the HRETIS method as an efficient alternative to enhance path decorrelation.

\subsection{Coarse-grained simulations}

\subsubsection{Slow permeation in the helper Hamiltonian increases exploration}

\begin{figure*}[!htb]
    \includegraphics[width=1\linewidth]{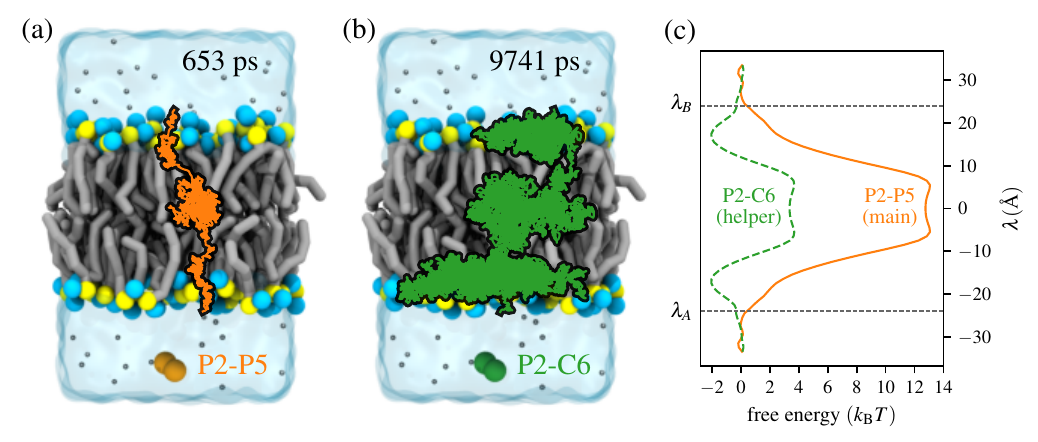}
    \caption{(a,b) The CG simulation boxes of permeant transport across a DPPC bilayer (visualized using MartiniGlass\cite{Brasnett2025MartiniGlass}):
    lipid tails (gray, licorice style), headgroup beads choline (light blue) and phosphate (yellow), ions (silver CPK spheres), and water (cyan surface). The permeants are beads P2-P5 (orange) or P2-C6 (green).
    The orange and green trajectories represent typical paths of each permeant in RETIS simulations.
    (c) Free energy profiles $F(\lambda)$ of the permeants P2-P5 and P2-C6 permeating through a DPPC membrane as a function of order parameter $\lambda$, obtained from RETIS.\cite{safaei2025exact,van2026generalized}
    Black dashed lines indicate $\lambda_A=-24~\text{\AA}$ and $\lambda_B=24~\text{\AA}$. }
	\label{Fig:CG_Box_FE}
\end{figure*}

To demonstrate the performance of HRETIS in a biologically relevant system, we performed coarse-grained (CG) simulations of permeant transport across a dipalmitoylphosphatidylcholine (DPPC) bilayer, inspired by our earlier all-atom MD path sampling study of 5-aminolevulinic acid (5-ALA) across a DPPC membrane \cite{safaei2025exact}. The Martini 3 classification \cite{souza2021martini} typically represents small drug-like molecules such as 5-ALA using three or more beads. In this work, a simplified representation of two beads per molecule was used to represent basic amino acids, in order to avoid additional complexities arising from angle and dihedral terms in the Martini force field.
Here, two bead combinations were chosen to explicitly mimic amino acid structures. The P2-P5 pair represents polar-polar interactions resembling glutamine (Fig.~\ref{Fig:CG_Box_FE}a), whereas the P2-C6 pair captures polar-apolar interactions characteristic of methionine (Fig.~\ref{Fig:CG_Box_FE}b). 
The main Hamiltonian $\mathcal{H}_\text{m}$ uses P2-P5, while the helper Hamiltonian $\mathcal{H}_\text{h}$ uses P2-C6.

The order parameter $\lambda$ for path sampling simulations was defined as the displacement along the membrane normal between the centers of geometry of the permeant and the membrane, $\lambda = z_\text{5-ALA}-z_\text{mem}$. To create a benchmark, RETIS was performed independently for the P2-P5 and P2-C6 systems, and the resulting trajectories were analyzed using a recently published reweighting scheme \cite{van2026generalized} to calculate the free energy profile $F(\lambda)$ as a function of the order parameter (Fig.~\ref{Fig:CG_Box_FE}c). 
The $F(\lambda)$ profiles in Fig.~\ref{Fig:CG_Box_FE}c indicate that P2-P5 encounters a higher free energy barrier during permeation ($\approx 13~k_BT$), whereas P2-C6 exhibits a lower barrier ($\approx 4~k_BT$). Both systems have a small metastable state at the membrane center. The P2-C6 system also has metastable states between the headgroup and tail regions which results in an average transitioning path length that is approximately 3.7 times longer than that of the P2-P5 system.

\begin{figure*}[!htb]
    \includegraphics[width=1\linewidth]{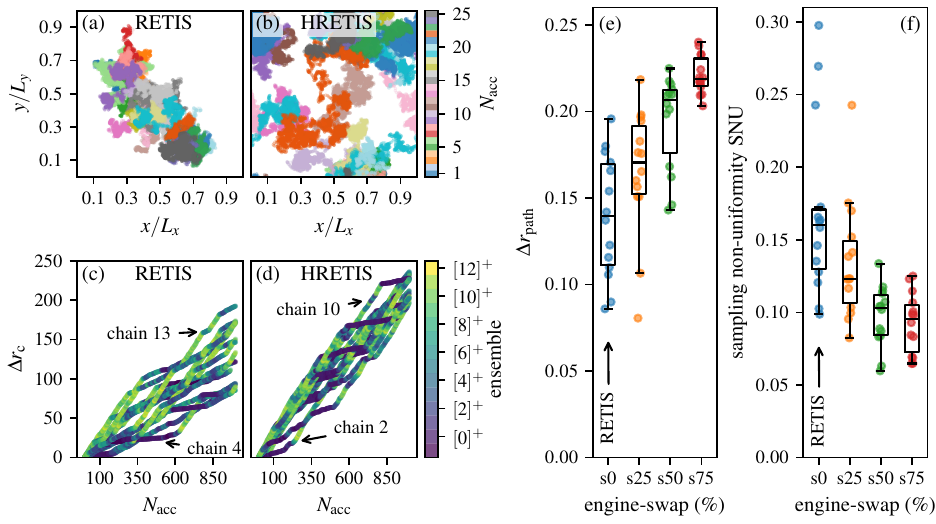}
    \caption{Assessment of phase space exploration in the $xy$-plane for RETIS and HRETIS of the P2-P5 permeant through DPPC membrane. (a,b) Trajectories for the first 25 accepted paths from a single chain (chain 5) projected onto the $xy$-plane for RETIS (left) and HRETIS 75\% (right). (c,d) Cumulative distance $\Delta r_\text{c}$ in the $xy$-plane between consecutive FCPs for all 14 chains, with color indicating the ensemble, for RETIS (left) and HRETIS 75\% (right). (e) Average distance $\Delta r_\text{path}$ in the $xy$-plane per path between consecutive FCPs, and (f) sampling non-uniformity $SNU$ in the $xy$-plane  
    as a function of engine-swap probability (0\% to 75\%) for all 14 MC chains.
    }
	\label{Fig:CG_analysis}
\end{figure*}

The long pathways arising from metastable states in the P2-C6 system lead to extensive exploration of phase space within the membrane. In contrast, most P2-P5 trajectories cross the membrane along a nearly straight pathway which indicates limited phase space exploration. Since new paths in RETIS are generated from existing ones, trajectories derived from the P2-P5 pathways are expected to remain localized in the same region of the $xy$-plane rather than exploring other regions of the membrane. Therefore, a relevant quantity to assess the improvement of the newly implemented HRETIS methodology relative to the RETIS approach is the extent of $xy$ exploration in each case. 

This lateral exploration of the permeant in the $xy$-plane may appear less significant here due to the homogeneous nature of the DPPC membrane, but even a homogeneous membrane has many slightly different permeation pathways because of local membrane fluctuations. The lateral exploration will hence contribute to faster path decorrelation in path sampling simulations. This effect is expected to be more significant in complex systems, such as membranes with mixed phospholipid compositions, inhomogeneous or phase-segregated membranes, or asymmetric membranes, where enhanced lateral sampling can substantially improve convergence. 

\subsubsection{Path decorrelation in the $xy$-plane}

Similarly to the 2D case of Section~\ref{sec:results2d}, here we also used the concept of path MC chains to study decorrelation among consecutively generated unique paths. Simulations were performed using HRETIS with engine-swap probabilities of 0\% (RETIS), 25\%, 50\%, and 75\% (s0, s25, s50, and s75). 
Figs.~\ref{Fig:CG_analysis}a-b show the full P2-P5 trajectories of the first 25 accepted paths in a random single MC chain (chain 5) for both RETIS and HRETIS (75\%). The trajectories are projected onto the $xy$-plane and show that HRETIS explores the $xy$-plane more efficiently, whereas in RETIS most paths remain confined to a relatively small region. In HRETIS, the helper Hamiltonian (P2-C6 permeant) enhances the path decorrelation in the $xy$-plane for the main Hamiltonian under study (P2-P5 permeant).

To quantify the enhancement in decorrelation, we again use the concept of the first crossing point (FCP) of a path in an ensemble, which was explained for the 2D system of Section~\ref{sec:results2d}. For each path's FCP, $(x^*,y^*)$ is used to denote the fractional $xy$-coordinates of the center of geometry of the permeant. These are obtained by normalizing the coordinates by the box length $L_x = L_y$, such that $0 \le x^*, y^* \le 1$.
Since defining distinct channels as in the previously studied 2D system is not straightforward for the CG system, the distance in the $xy$-plane using the minimal image convention is calculated between the 
$(x^*,y^*)$ positions of consecutive FCP's in a chain. The cumulative distance $\Delta r_{\text{c}}$ of a chain of $N_\text{acc}$ paths is then
\begin{equation}
\Delta r_{\text{c}} = \sum_{k=1}^{N_\text{acc}-1} \sqrt{(x_{k+1}^* - x_k^*)^2 + (y_{k+1}^* - y_k^*)^2}
\end{equation}
where the sum runs over all paths in the chain.
Figs.~\ref{Fig:CG_analysis}c-d show $\Delta r_{\text{c}}$ of P2-P5 for the first 1000 accepted paths across all 14 MC chains, with colors indicating the corresponding ensembles. It clearly shows that the distance traveled per path in HRETIS is higher than that in RETIS, confirming that HRETIS enhances path decorrelation for the P2-P5 permeant.

In the $\infty$RETIS framework, a single path can belong to different ensembles in subsequent MC cycles \cite{zhang2024highly}. 
An important observation in Figs.~\ref{Fig:CG_analysis}c-d is that the paths in MC chains can remain confined to lower ensembles (dark blue) for extended periods before being swapped to higher ones (green, yellow). This behavior is inherent to the infinite-swap of $\infty$RETIS, since two chains are always sampling the $[0^-]$ and $[0^+]$ ensembles. This effect is more pronounced in RETIS chains (e.g.\ chain 4 in Fig.~\ref{Fig:CG_analysis}c), where the first two ensembles $[0^-]$ and $[0^+]$ rely solely on shooting move and do not benefit from the decorrelation provided by the wire-fencing move. In contrast, in HRETIS, paths in the $[0^-]$ and $[0^+]$ ensembles also benefit from Hamiltonian exchange (e.g.\ chain 2 in Fig.~\ref{Fig:CG_analysis}d). 
Overall, when comparing all 14 MC chains across the four simulation settings (s0 to s75) in Fig.~\ref{Fig:CG_analysis}e, the distance $\Delta r_{\text{path}}$ traveled per path in the $xy$-plane,
\begin{equation}
\Delta r_{\text{path}} =
\frac{\Delta r_{\mathrm{c}}}{N_\text{acc}-1},
\end{equation}
increases with the engine-swap probability in HRETIS and, in all cases, exceeds that of the RETIS methodology (i.e.\ HRETIS 0\%).

Fig.~\ref{Fig:CG_analysis}f further quantifies the uniformity of sampling in the $xy$-plane. For each chain, all accepted trajectories are projected onto the $xy$-plane and binned into a two-dimensional histogram with 100x100 bins without MC weights. Since the membrane is homogeneous, no spatial preference is expected, and ideal sampling would yield a flat histogram even without MC weights. To quantify how well path sampling matches this ideal behavior, the sampling non-uniformity $SNU$ is defined as the ratio of the standard deviation of the bin counts to their mean:
\begin{equation}
SNU = \frac{\sigma_{\mathrm{bins}}}{\mu_{\mathrm{bins}}}
\end{equation}
where $\sigma_{\mathrm{bins}}$ and $\mu_{\mathrm{bins}}$ denote the standard deviation and mean of the histogram bin counts, respectively, and higher values of $SNU$ indicate a less uniform distribution. 

The $SNU$ values of the 14 MC chains of the P2-P5 permeant in Fig.~\ref{Fig:CG_analysis}f
show a clear trend in which increasing the engine-swap probability setting (s0 to s75) leads to more homogeneous sampling in the $xy$-plane. In addition, the outliers observed in the RETIS methodology (s0)
indicate that three chains (chains 1, 2, and 5) become trapped in localized spatial regions. In contrast, this number reduces to one (chain 14) for HRETIS with 25\% engine-swap probability (s25) and to zero for higher swapping percentages (s50, s75), confirming that the Hamiltonian exchange move reduces the risk of being trapped in a part of phase space.

\subsubsection{HRETIS converges faster than RETIS}

The phase space exploration already demonstrates the advantage of HRETIS over RETIS. This increased phase space exploration is also suspected to be reflected in the convergence of the total crossing probability $P_A(\lambda_B|\lambda_A)$ for the P2-P5 permeant. To obtain reliable error estimates for the path sampling simulations, five runs were performed for each method (RETIS and HRETIS) with all simulations initiated from the same initial set of paths. The first 10,000 accepted paths of each run were excluded from the analyses to eliminate the influence of the common initial path set. 
In addition, to compare the convergence of the methodologies,\cite{kang2026convergence} a reliable reference value for the crossing probabilities is required. Since the probability of crossing the first few interfaces is sufficiently high, these local crossing probabilities can also be calculated from standard equilibrium MD simulations.
Therefore, ten independent MD simulations of 1 microsecond each were performed to provide a reference value for the comparison of the local crossing probability $P_A(\lambda_{i+1}|\lambda_i)$ convergence in the lower ensembles. This was done for both the P2-P5 and P2-C6 permeant.

\begin{figure}[!htb]
    \includegraphics[width=1\linewidth]{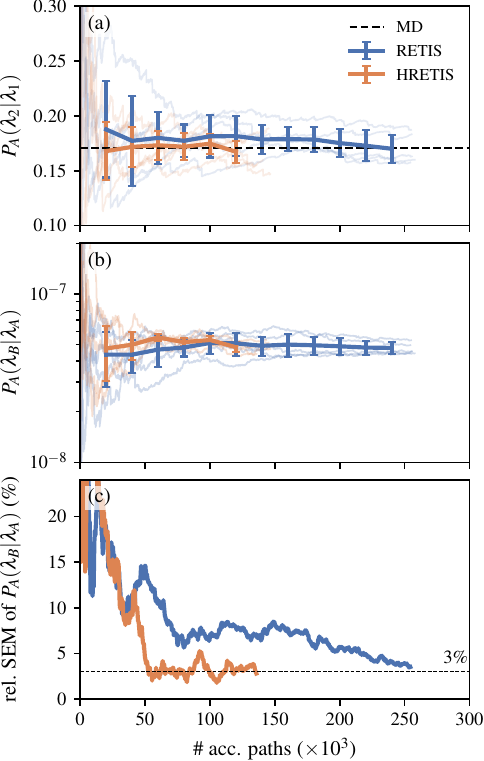}
    \caption{Permeation of P2-P5 permeant through DPPC membrane. Assessing the convergence of the running estimate of crossing probability as a function of the number of accepted paths $N_\text{acc}$ for the P2-P5 Hamiltonian. (a) Comparison of the local crossing probability $P_A(\lambda_2 | \lambda_1)$ for ensemble $[1^+]$ in MD, RETIS, and HRETIS. (b) Comparison of the total crossing probability $P_A(\lambda_B | \lambda_A)$ in RETIS and HRETIS. (c) Relative standard error of the mean (rel.\ SEM) of $P_A(\lambda_B | \lambda_A)$ computed over five runs for RETIS and HRETIS. Five path sampling runs are shown for each methodology as transparent lines, while solid lines indicate the mean values. The error bars in (a-b) are $\pm 2$ SEM over the five runs.}
	\label{Fig:pcross}
\end{figure}

Fig.~\ref{Fig:pcross}a compares the running estimate of the local crossing probability $P_A(\lambda_2 | \lambda_1)$ under $\mathcal{H}_\text{m}$ evaluated from ensemble $[1^+]$ between the MD reference simulations, RETIS, and HRETIS with a 75\% engine-swap probability as a function of the number of accepted paths $N_\text{acc}$. 
The mean values over the runs indicate that HRETIS converges to the reference value significantly faster than RETIS. 
Similar plots for the other lower ensembles are presented in Supplementary Fig.~S11, where the same trend is observed for $[2^+]$ and $[3^+]$. 
However, for $[0^+]$, both RETIS and HRETIS slightly underestimate the local crossing probability $P_A(\lambda_1 | \lambda_0)$. 

The total crossing probability $P_A(\lambda_B | \lambda_A)$ for five runs of both RETIS and HRETIS simulations (Fig.~\ref{Fig:pcross}b) clearly shows that the variation between runs can be substantial and unpredictable, although it is smaller for HRETIS than for RETIS. The spread among the runs confirms that assessing the convergence of the total crossing probability from extended path sampling simulations is not straightforward. Fig.~\ref{Fig:pcross}c presents the standard error of the mean (SEM) computed from the five runs as a function of $N_\text{acc}$. HRETIS consistently exhibits lower errors than RETIS and, more importantly, reaches errors below 3\% after approximately 50,000 accepted paths, whereas RETIS does not achieve the same level of accuracy even after 250,000 accepted paths in each of the five runs. Interestingly, this trend is also observed for the crossing probability of the helper Hamiltonian $\mathcal{H}_\text{h}$ (Supplementary Fig.~S12), where HRETIS reaches the same level of accuracy approximately two times faster than RETIS. This is especially significant given that HRETIS achieves this performance despite using a simple MC weighting scheme for the helper Hamiltonian, whereas the RETIS simulation is aided by the infinite-swap approach. For the 2D MM0 potential, the convergence of $P_A(\lambda_B | \lambda_A)$ follows a similar trend, where HRETIS reaches errors below 1.5\% approximately twice as fast as RETIS (Supplementary Fig.~S9).

Permeability is another quantity that can be obtained from RETIS crossing probabilities, and it captures both kinetic and thermodynamic aspects of membrane permeation. In Ref.~\citenum{ghysels2021exact}, it was shown that the permeability $P$ can be obtained from the RETIS crossing probability $P_A(\lambda_B|\lambda_A)$. The procedure for converting this probability into a permeability (with units of length per time) is detailed in the Supplementary Information (Supplementary Fig.~S13). The permeability was computed for both 
both the main and helper Hamiltonians,
where good agreement is observed between RETIS and HRETIS. However, kinetics from CG simulations without appropriate correction cannot be directly interpreted as physical time scales. This is due to the reduced number of degrees of freedom and the reduced effective bead friction arising from the smoother energy landscape~\cite{fritz2011multiscale, rudzinski2019recent, sadeghi2020large}.

\subsubsection{Computational efficiency}

It should be noted that using the number of accepted paths $N_\text{acc}$ can be misleading when comparing RETIS and HRETIS from a computational cost perspective. This arises for two reasons. First, for every accepted path generated in the main Hamiltonian, an additional accepted path is generated in the helper Hamiltonian in HRETIS, which increases the computational cost depending on how expensive the helper Hamiltonian is. Second, HRETIS shows a higher total number of rejections, which occur at three stages in the engine-swap move: the $\Delta\Delta U$ criterion, path generation in the first Hamiltonian, and path generation in the second Hamiltonian.

Early rejection via $P^{\Delta\Delta U}_\text{acc}$ (Eq.~\ref{eq:acc-criterion-DDU}) at the $\Delta\Delta U$ level is almost costless, since only two phase points are evaluated using the MD engine to compute this quantity. Rejections that occur after $\Delta\Delta U$ acceptance, during MD propagation in the first Hamiltonian, do not introduce additional cost compared to a standard shooting move (i.e.\ RETIS). In contrast, rejections that occur during MD propagation in the second Hamiltonian are costly. However, they occur rarely, as in the wire-fencing scheme with high-acceptance, the acceptance ratio is expected to be higher than 90\%.

Similarly as for the 2D system, the computational cost was measured as the number of force evaluations required per accepted path.
Overall, the cost of HRETIS is higher than that of the RETIS simulation (Supplementary Fig.~S14.).
It remains comparable to RETIS at low engine-swap probabilities and gradually increases with increasing engine-swap probability. This additional cost would be reduced or negligible when the helper Hamiltonian is significantly less expensive to evaluate. Furthermore, when statistics from both Hamiltonians are of interest, the effective cost of HRETIS decreases, since each engine-swap move generates two accepted paths per MC cycle, one in each Hamiltonian, whereas RETIS produces only a single accepted path per cycle.

\section{Discussion \label{Sec:Discussion}}

We extended the existing path sampling methodology RETIS to HRETIS by incorporating Hamiltonian replica exchange into the framework. This allows configurations to be exchanged between two Hamiltonians, where a helper Hamiltonian $\mathcal{H}_\text{h}$ that explores phase space more efficiently assists the main Hamiltonian $\mathcal{H}_\text{m}$. This concept is especially attractive when the main Hamiltonian cannot adequately explore phase space on its own in a feasible amount of simulation time. The engine-swap move therefore accelerates convergence in path sampling simulations. We validated the new algorithm using 1D potentials, demonstrated the decorrelation with 2D model systems, and applied it to high-dimensional CG simulations.

HRETIS was able to efficiently sample the phase space in the 2D models, where two distinct channels connected the stable states. This was assessed by monitoring 
the coordinate $y$ orthogonal to the order parameter and counting
how frequently consecutively accepted paths switched between the two channels.
A lack of switching in standard RETIS MC chains indicated that the simulation becomes trapped in a single channel, which is equivalent to insufficient sampling of orthogonal degrees of freedom due to the presence of orthogonal barriers in the collective variable space (Fig.~\ref{Fig:2d_analysis}). The ensuing lack of path decorrelation in RETIS is a common issue in path sampling simulations and often arises from a suboptimal choice of the order parameter.

HRETIS was also able to demonstrate its capabilities for path decorrelation
in a more complex CG system of different amino acids permeating a homogeneous DPPC membrane. This was assessed by examining 
how well the sampled permeation pathways were spread over the cross section of the membrane.
The need for lateral exploration
is expected to become even more of a bottleneck in complex heterogeneous environments, such as membranes with mixed phospholipid compositions, inhomogeneous or phase-segregated membranes, or asymmetric membranes, where enhanced lateral sampling could significantly improve convergence.

In HRETIS, the helper Hamiltonian $\mathcal{H}_\text{h}$ is not the primary target of the calculation, since its dynamics are typically either computationally inexpensive or less accurate than those of the main Hamiltonian $\mathcal{H}_\text{m}$. One example is a drug molecule permeating a membrane, where a Drude force field could be employed as the main Hamiltonian to obtain more accurate kinetics for a polarizable drug molecule at a high computational cost, while a standard CHARMM force field~\cite{klauda2010update} is used as the helper Hamiltonian with lower accuracy and lower cost. Another example is a system represented at all-atom resolution as the main Hamiltonian and by a CG model as the helper Hamiltonian. In this case, the CG kinetics may not be physically meaningful, but the model can still exhibit substantially different dynamics that enhance phase space exploration. In such cases, the computational cost of the helper Hamiltonian is orders of magnitude lower than that of the main Hamiltonian, so the additional cost introduced by the helper in HRETIS would be negligible in practice. Another powerful interpretation of HRETIS could be temperature-based replica exchange, as it can be viewed in a similar Hamiltonian exchange framework, where temperature is effectively absorbed into the Hamiltonian through an energy rescaling. Similarly as in replica exchange MD,\cite{abrams2013enhanced} the higher temperature replicas could act as helper Hamiltonians that enhance exploration of the main Hamiltonian.

Nevertheless, in many cases the kinetics of both Hamiltonians are of interest. One example is the membrane permeation of structurally related drug molecules. A relevant case is the permeation of 5-ALA through a DPPC membrane, which was recently studied using RETIS and required several weeks of path sampling simulations to obtain converged kinetics. Such studies could potentially benefit from HRETIS if the system were paired with a closely related molecule, such as a methylated derivative of 5-ALA, which is used for similar drug treatment applications but exhibits different permeation kinetics. In this setup, each system could benefit from the sampling of the other which allows HRETIS to improve convergence while simultaneously providing kinetic information for both drug molecules.

As for the choice of the main and helper Hamiltonians, in the current implementation of HRETIS, the statistics extracted from the main Hamiltonian $\mathcal{H}_\text{h}$ are expected to be more accurate since the infinite-swap scheme is implemented only for $\mathcal{H}_\text{m}$, whereas $\mathcal{H}_\text{h}$ uses simple MC weighting. In addition, the number of paths generated for $\mathcal{H}_\text{h}$ per accepted MC cycle decreases as the HRETIS engine-swap frequency decreases, while one path is always generated for $\mathcal{H}_\text{m}$ in each accepted MC cycle, independent of the engine-swap frequency. Still, it is conceptually possible to implement more advanced helper exploration in the future if needed, e.g.\ when working with temperature-based Hamiltonian exchange in HRETIS.

There is a subtle technical point when both Hamiltonians have the same level of complexity and therefore comparable computational cost. In such cases, the usual distinction between low-cost and high-cost Hamiltonians becomes irrelevant. As observed for CG amino acid permeation in the present study, the average path length associated with one Hamiltonian can differ significantly from that of the other, potentially due to the presence of metastable states. This can increase the computational cost of computing the kinetics of the main Hamiltonian with HRETIS, depending on how slow the helper Hamiltonian is. However, it should be noted that in such systems, the statistics for both Hamiltonians are of interest, and the computational effort is effectively used to generate paths for both Hamiltonians simultaneously.

Moreover, it may appear counterintuitive that the CG system with longer paths (P2-C6 permeant) could assist the one with shorter paths (P2-P5 permeant). One could hypothesize that the roles should be swapped, i.e.\ the system without metastable states (shorter path lengths) assists the system with metastable regions and longer paths by reducing the effective time spent in those regions. However, on second thought, the helper Hamiltonian does not drastically affect the path length distribution of the main Hamiltonian since detailed balance respects the true path distributions,
and phase space exploration does not improve under this setup.
Therefore, the role of the helper Hamiltonian was given to the P2-C6 system with the long paths, and this leads to the conclusion that longer paths generally correspond to more extensive exploration of configuration space and therefore sample the phase space more effectively than shorter paths.

\section{Methods}

\subsection{RETIS concept \label{sec:retis-review}}

Consider a system with two stable states A and B separated along a reaction coordinate $\lambda(x)$, which maps a phase point $x$ onto a scalar order parameter describing progress along the transition pathway. The states are defined by two boundaries $\lambda_A$ and $\lambda_B$, with $\lambda_A < \lambda_B$. A configuration is assigned to state A when $\lambda(x) < \lambda_A$ and to state B when $\lambda(x) > \lambda_B$. The probability $P_A(\lambda_B | \lambda_A)$ of reaching state B before returning to state A, given that trajectories are initiated in A, can be expressed as an exact factorization
\begin{equation}  
P_A(\lambda_B | \lambda_A) = \prod_{i=0}^{n-1} P_A(\lambda_{i+1} | \lambda_i), 
\label{eq:total_pcross}
\end{equation}
where $P_A(\lambda_{i+1} | \lambda_i)$ denotes the local crossing probability that a trajectory crossing interface $\lambda_i$, having originated in state A, reaches the next interface $\lambda_{i+1}$ before returning to A. These local crossing probabilities are typically significantly larger than the total crossing probability, which makes their estimation via MC sampling feasible.

In practice, this is achieved by sampling the $[i^+]$ path ensemble, which consists of trajectories that start in state A, cross interface $\lambda_i$, and reached either state A or B. After sufficiently long MC sampling, $P_A(\lambda_{i+1} | \lambda_i)$ is estimated as the fraction of trajectories in the $[i^+]$ ensemble that reach $\lambda_{i+1}$ before returning to A. 
These estimates are then combined by path reweighting \cite{rogal2010reweighted, van2016analyzing} based on the weighted histogram analysis method (WHAM) \cite{ferrenberg1989optimized, kumar1992weighted, roux1995calculation} to obtain $P_A(\lambda_B | \lambda_A)$.

\subsection{HRETIS improvements \label{sec:improvements}}

The HRETIS algorithm was implemented in the \texttt{infRETIS} software package~\cite{infretissoftware}. The \texttt{infRETIS} code is interfaced with multiple MD engines such as the Atomic Simulation Environment (ASE) \cite{hjorth2017atomic} and GROMACS \cite{GROMACS}. 
The current implementation includes several algorithmic improvements (Supplementary Fig.~S2).\cite{hretis_git}
\begin{enumerate}
\item When wire-fencing is used, the shooting point selection follows the wire-fencing scheme,\cite{zhang2023enhanced} in which phase points are not selected randomly from the entire path but only from points above the interface $\lambda_i$ in each ensemble.
\item The Hamiltonian exchange algorithm is explained for a randomly selected ensemble [$i^+$] in Section~\ref{sec:HRETIS_algorithm}. The engine-swap has also been implemented in the $[0^-]$ ensemble which covers the region of reactant state A ($\lambda<\lambda_A$).
\item The helper exploration move is introduced for the helper Hamiltonian $\mathcal{H}_\text{h}$. This allows the paths in the helper Hamiltonian to explore their phase space independently of the main Hamiltonian with a certain probability, rather than only receiving phase points from the main Hamiltonian.
\item For the main Hamiltonian, the infinite-swap scheme used in $\infty$RETIS methodology is applied every time a new valid RETIS path is created. For the helper Hamiltonian, path weights are updated using simple MC rules. If a path is rejected, it is preserved and its weight is incremented by one. If it is accepted, the weight remains unchanged and the new trajectory replaces the old one.
\item The factors in $P'_\text{acc}$ (Eq.~\ref{eq:acc-criterion-method2}) are evaluated in a strategic order. Following the evaluation of $P^{\Delta\Delta U}_\text{acc}$ (Eq.~\ref{eq:acc-criterion-DDU}) and the first possible early rejection, a path is generated and evaluated using the helper Hamiltonian $\mathcal{H}_\text{h}$ first, assuming propagation with $\mathcal{H}_\text{h}$ is computationally less expensive. This introduces an additional early rejection during HRETIS propagation. Consequently, unnecessary propagation with the more expensive $\mathcal{H}_\text{m}$ can be avoided. In the future, the order of $\mathcal{H}_\text{m}$ and $\mathcal{H}_\text{h}$ propagation could be implemented as a user setting to cover the case of the main Hamiltonian being the least expensive.
\end{enumerate}

Another algorithmic aspect, although it should not necessarily be regarded as as improvement, concerns the treatment of velocities following a Hamiltonian exchange in step 3 and before propagation in step 4 of Section~\ref{sec:HRETIS_algorithm}. After exchanging phase points, the velocities are redrawn from the Maxwell-Boltzmann distribution rather than reusing the old velocities, similarly to aimless shooting \cite{peters2006obtaining, mullen2015easy, falkner2025revisiting}. This approach was employed to maintain consistency with the existing \texttt{infRETIS} implementation. 
Incorporation of the Maxwell-Boltzmann distribution into the generation probability in the engine-swap move (Eq.~\ref{eq:Pgen}) leads to the same acceptance criterion as in Eq.~\ref{eq:acc-criterion-DDU}. Detailed balance is thus also be achieved when the exchanged phase points with the newly redrawn velocities are used.

\subsection{One-dimensional systems \label{sec:1dsettings}}

A 1D Langevin particle was simulated with mass $m = 39.948$ amu and friction constant $\gamma = 0.003~\text{fs}^{-1}$, mimicking an argon atom. The particle was simulated with ASE\cite{hjorth2017atomic} at a temperature $T = 300$ K using an MD time step of $2$ fs. 

The potential function $u(x)$ is a simple cosine-shaped bump, similar to earlier work,\cite{ghysels2021exact, vervust2026estimating}
\begin{equation}\label{eq:u1d}
u(x)=
\frac{V_0}{2}\left[\cos\!\left(\pi (x-s)\right)+1\right]
\end{equation}
for  $|x-s|\leq 1$ and 0 elsewhere,
where $V_0$ and $s$ control the barrier height and shift, respectively. The values of $V_0$ were 0, 1, 1.5, and 1~$k_{\text{B}}T$ for the flat, bump, high-bump, and shift-bump profiles, respectively (see Fig.~\ref{Fig:1d_potentials}). The shift parameter $s$ was set to 0 for all profiles except shift-bump, for which $s = -0.5$~\AA.

In the RETIS and HRETIS simulations, the order parameter $\lambda = x$ was used, with $\lambda_A = -2~\text{\AA}$ and $\lambda_B = 2~\text{\AA}$. The intermediate interfaces were placed at $-1$, 0, and 1 (in {\AA}), and an additional $\lambda_{-1}$ interface, required for unbounded systems~\cite{ghysels2021exact}, was set at $-3~\text{\AA}$.

All path sampling simulations comprised of 100,000 MC steps. The first $N_\text{skip}= 5000$ accepted paths were discarded, and the subsequent 25,000 accepted paths were used for the analysis. 

For the $[0^+]$ and $[0^-]$ ensembles, each MC cycle consisted of the standard shooting move. 
For all other ensembles $[i^+]$, the wire-fencing move was used with four subtrajectories and the high-acceptance scheme~\cite{riccardi2017fast, zhang2023enhanced}. The subcycle parameter, which controls how often $\lambda$ is evaluated, was set to 1, meaning that $\lambda$ was evaluated at every MD step. In the HRETIS simulations, the engine-swap probability and the helper exploration probability were both set to 50\%. 

Error bars represent the standard error of $P_A(\lambda_B|\lambda_A)$, estimated using recursive block error analysis, with a minimum of five blocks to ensure statistical reliability \cite{vervust2024pyretis}. 
The final reported error is computed as the average over the second half of the considered block lengths. \cite{vervust2025path, safaei2025exact}

\subsection{Two-dimensional systems \label{sec:2dsettings}}

The same argon-like particle of Section~\ref{sec:1dsettings} was modeled to study transport across a membrane, with the same mass, friction constant, temperature, and Langevin integrator settings.

A 2D potential energy surface $V(y,z)$ was taken to model a membrane with two distinct permeation pathways, similar to earlier work,\cite{ghysels2021exact}
\begin{equation}
\begin{aligned}
V(y,z) &= e^{-cz^2} 
\left( V_1 + A + A \sin \frac{2 \pi y}{L_y} \right. \\
       & \left. + B + B \cos \frac{4 \pi y}{L_y}
       \right), \\
 A &= \frac{V_{2} - V_{1}}{2} \\
 B &= \frac{V_{\mathrm{max}}}{2} - \frac{V_1}{4} - \frac{V_2}{4}
\end{aligned}   
\end{equation}
where $y$ is periodic with period $L_y$.
The settings were $c = 1/$\AA$^2$ and $L_y = 6$\,{\AA},
while
$V_{1}$, $V_{2}$, and $V_\text{max}$ 
are given in Table~\ref{tab:2dpotentials}.

\begin{table}[tbh]
\centering
\begin{tabular}{llccc}
\hline
2D-potential & & $V_{1}$ & $V_{2}$ & $V_\text{max}$ \\
\hline
main $\mathcal{H}_\text{m}$ & MM0 & 10   & 11 & 20   \\
\hline
helper $\mathcal{H}_\text{h}$\\
flat membrane     & FM  & 0 & 0 & 0  \\
model membrane 1  & MM1 & 0 & 1   & 40   \\
model membrane 2  & MM2 & 5 & 5.5 & 20 \\
\hline
\end{tabular}
\caption{Parameters defining the 2D-potentials $V(y,z)$ for a model membrane. In units $k_\mathrm{B}T$.}
\label{tab:2dpotentials}
\end{table}

The $z$-coordinate describes the permeation and is used as the order parameter $\lambda$.
The set of interfaces included 12 $\lambda_{i}$ values located at $z = -1.5$, -1.3, -1.15, -1.0, -0.9, -0.8, -0.7, -0.6, -0.5, -0.4, -0.2, \text{and} 1.2 \si{\angstrom}, while the additional $\lambda_{-1}$ interface\cite{ghysels2021exact} was positioned at $z = -4.5$ \si{\angstrom}.

All path sampling simulations were performed for 100,000 MC cycles using different move types: shooting (sh), wire-fencing with one subtrajectory (wf1), or wire-fencing with four subtrajectories (wf4). No paths were skipped in the reported results of Section~\ref{sec:results2d}.
Reference RETIS simulations were performed using only the membrane potential (MM0). 
HRETIS simulations used the membrane model MM0 as the main Hamiltonian $\mathcal{H}_\text{m}$ in combination with another model membrane (FM, MM1, or MM2) as helper Hamiltonian $\mathcal{H}_\text{h}$.

Two settings were varied to study their impact: the probability of performing an engine-swap move (label s: 25\%, 50\%, 75\%, or 100\%) and, when not performing an engine-swap, the probability to perform a helper exploration move in the helper Hamiltonian (label e: 0\%, 50\%, or 100\%).

\subsection{Coarse-grained simulations \label{sec:cgsettings}}

The CG system was built with the CHARMM-GUI web server~\cite{jo2008charmm}, where a homogeneous symmetric bilayer containing 100 DPPC molecules was placed in a tetragonal simulation box of approximately 5.66~nm $\times$ 5.66~nm $\times$ 9.20~nm with periodic boundary conditions. The system contained 1361 water beads, 16 Na$^+$ beads, and 16 Cl$^-$ beads. The Martini~3 force field \cite{souza2021martini} was used for the bead interactions. One permeant molecule was added, consisting of two Martini beads: either P2-P5 or P2-C6.
Non-bonded interactions were calculated using the Verlet cutoff scheme with a cutoff of 1.1~nm for both Lennard-Jones and Coulomb interactions. Electrostatic interactions were treated using the reaction field method with a relative dielectric constant of \(\epsilon_r = 15\). Van der Waals interactions were truncated at 1.1~nm using a potential shift modifier. 
All MD simulations were performed with GROMACS 2024.4~\cite{GROMACS} with an integration time step of 20~fs. The temperature was controlled at 323~K using the velocity rescale thermostat with a coupling constant of 1~ps~\cite{bussi2007canonical}. This temperature is above the gel to liquid phase transition temperature of DPPC~\cite{tong2019experimental}. The pressure was maintained at 1~bar using the stochastic cell rescaling barostat in a semi-isotropic mode, with a time constant of 12~ps and a compressibility of \(3 \times 10^{-4}\)~bar$^{-1}$~\cite{bernetti2020pressure}. Each system was equilibrated for 500~ns in the NPT ensemble and the final configuration was used as the starting point for path sampling simulations. All subsequent path sampling simulations were performed in the NPT ensemble using the same settings.

To start the path sampling simulations, a set of interfaces and of initial paths were required. These were generated using the \emph{inf-init} module of the infRETIS package.
This module optimizes the intermediate interfaces $\lambda_i$ between $\lambda_A$ and $\lambda_B$ 
based on the available hardware, while it ensures a local crossing probability $P_A(\lambda_{i+1}|\lambda_i)$ of at least 0.3 for each ensemble $[i^+]$. This optimization was performed over 5 iterations with 50, 100, 250, 500, 2000, and 4000 MC cycles. The final output of the module consisted of 12 interfaces $\lambda_i$ placed between $\lambda_0=\lambda_A=-24~\text{\AA}$ and $\lambda_{13}=\lambda_B=24~\text{\AA}$, which led to a total of 14 ensembles with initial paths. An additional interface at $\lambda_{-1}=-34$\,{\AA} was introduced to prevent crossings through the periodic boundary. The initial paths generated by \emph{inf-init} for the 14 ensembles were used to initiate all subsequent path sampling simulations. A subcycle value of 10 was employed for all simulations, meaning that $\lambda$ is evaluated every 0.2\,ps.

Two sets of path sampling simulations were performed. The first set included 
a RETIS simulation for the main Hamiltonian (P2-P5), a RETIS simulation for the helper Hamiltonian (P2-C6), and three HRETIS simulations with engine-swap frequencies of 25, 50, and 75\% (Supplementary Table S2). 
Each simulation comprised 30,000 MC cycles. These simulations were used to generate Figs.~\ref{Fig:CG_Box_FE}-\ref{Fig:CG_analysis}.

The second set of simulations included a RETIS simulation for the main Hamiltonian (P2-P5), a RETIS simulation for the helper Hamiltonian (P2-C6), and an HRETIS simulation with engine-swap frequency of 75\% (Supplementary Table S3). For each case, five runs were performed for 300,000 MC cycles to estimate statistical uncertainties and assess convergence. The simulations were used in Fig.~\ref{Fig:pcross}.
The first $N_\text{skip}=10,000$ accepted paths were discarded in the analysis.

Ten independent equilibrium MD simulations were performed from the same initial configuration, each 1 microsecond (10 microseconds in total)
to calculate the local crossing probabilities for the first few ensembles 
for the main (P2-P5) and helper (P2-C6) Hamiltonian.
These simulations were used for Fig.~\ref{Fig:pcross}.

\appendix

\section{HRETIS acceptance criterion \label{sec:DetailedBalance}}

Detailed balance for the Hamiltonian exchange move is derived by considering the joint
state $z = (X_\text{m}, X_\text{h})$ of the two systems at once \cite{van2005elaborating}, when $X_\text{m}$ is the current path for the main Hamiltonian $\mathcal{H}_\text{m}$ and $X_\text{h}$ for the helper Hamiltonian $\mathcal{H}_\text{h}$.
An exchange of Hamiltonians causes a hop in the Markov chain from the old state $z^\text{(o)} = (X_\text{m}^\text{(o)}, X_\text{h}^\text{(o)})$ to the new state $z^\text{(n)} = (X_\text{m}^\text{(n)}, X_\text{h}^\text{(n)})$. 
The detailed balance equation is
\begin{equation}
P(z^{\text{(o)}}) 
P(z^{\text{(o)}}\rightarrow z^{\text{(n)}})
=
P(z^{\text{(n)}}) 
P(z^{\text{(n)}}\rightarrow z^{\text{(o)}})
\end{equation}
where the probability to go from state $z^{\text{(o)}}$ to state $z^{\text{(n)}}$ is the product of the generation probability and acceptance probability,
\begin{equation}
      P(z^{\text{(o)}} \rightarrow z^{\text{(n)}}) 
    = P_{\text{gen}}(z^{\text{(o)}} \rightarrow z^{\text{(n)}})
      P_{\text{acc}}(z^{\text{(o)}} \rightarrow z^{\text{(n)}})
\end{equation}
To satisfy detailed balance, the acceptance probabilities must therefore obey
\begin{equation}
    \frac{P_\text{acc}(z^{\text{(o)}}\rightarrow z^{\text{(n)}})}{P_\text{acc}(z^{\text{(n)}}\rightarrow z^{\text{(o)}})} 
     = \frac{P(z^{\text{(n)}})}{P(z^{\text{(o)}})}
     \frac{P_{\text{gen}}(z^{\text{(n)}}\rightarrow z^{\text{(o)}})}{P_{\text{gen}}(z^{\text{(o)}}\rightarrow z^{\text{(n)}})}
    \label{eq:ratioPacc}
\end{equation}

Let us first compute the path probability $p_\text{m}[X]$ to observe path $X=(x_0,\ldots,x_{N})$ of length $N+1$ under the main Hamiltonian $\mathcal{H}_\text{m}$.
If the system has microscopically time-reversible dynamics, we can write the path probability as
\begin{align}
    p_\text{m}[X]
    = p_\text{m}(x^k)
    & \prod_{j=k}^{N-1} p_\text{m}(x^{j} \rightarrow x^{j+1})  \nonumber\\
    & \times \prod_{j=1}^{k} p_\text{m}(\bar{x}^{j} \rightarrow \bar{x}^{j-1})
\label{eq:path-density}
\end{align}
where $x^{k}$ is the $k^{th}$ phase point in path $X$.
Here, $\bar{x}=(r,-p)$ refers to the momenta-reversed phase point: if $x = (r, p)$ with $r$ the configuration and $p$ the momenta of all particles, then $\bar{x} = (r, -p)$. 
The propagator dynamics $p_\text{m}(x^{j}\rightarrow x^{j+1})$ depend on $\mathcal{H}_\text{m}$.
In Eq.~\ref{eq:path-density}, the probability $p_\text{m}(x^k)$ of observing phase point $x^k$ under the main Hamiltonian refers to the Boltzmann equilibrium distribution,
\begin{equation}
    p_\text{m}(x) = \exp(-\beta \mathcal{H}_\text{m}(x))/Q_\text{m}
\label{eq:phasepoint-density}
\end{equation}
with $Q_\text{m}=\int dx\,\exp(-\beta \mathcal{H}_\text{m}(x))$ the normalizing constant.

Within the $[i^+]$ ensemble, the path distribution is slightly rebalanced and becomes $p_\text{m}^{\lambda_i}[X]$,
\begin{equation}
    p_\text{m}^{\lambda_i}[X]
    =\frac{1}{Z_{\text{m}}^{\lambda_i}} \mathds{1}_{i}[X]p_\text{m}[X]
    \label{eq:path-density-lambda}
\end{equation}
with $Z_{\text{m}}^{\lambda_i} = \int \mathcal{D}X \mathds{1}_{i}[X]p_\text{m}[X]$
the normalizing factor and the indicator function $\mathds{1}_{i}[.]$.

For the helper Hamiltonian $\mathcal{H}_\text{h}$, similar relations hold for $p_\text{h}[X]$, $p_\text{h}(x)$, and $p^{\lambda_i}_\text{h}[X]$.
This leads to the joint probability distribution $P(z)$,
\begin{equation}
    P(z) = p_\text{m}^{\lambda_i}[X_{\text{m}}]
    \,     p_\text{h}^{\lambda_i}[X_{\text{h}}]
    \label{eq:Pz}
\end{equation}
in which Eq.~\ref{eq:path-density-lambda}, Eq.~\ref{eq:path-density}, and Eq.~\ref{eq:phasepoint-density} (and similar equations for the helper Hamiltonian) can be substituted.

The final quantity needed to compute the ratio of acceptance probabilities in Eq.~\ref{eq:ratioPacc} is the generation probability $P_{\text{gen}}(z^{\text{(o)}}\rightarrow z^{\text{(n)}})$.
First, a phase point $x_{\text{m}}^\dagger$ and point $x_{\text{h}}^\dagger$ 
are randomly selected from the old paths, that have $N_{\text{m}}^{\text{(o)}}$ and $N_{\text{h}}^{\text{(o)}}$ phase points.
The selection probabilities are
\begin{subequations}
    \begin{align}
        P_\text{sel}[X_{\text{m}}^{\text{(o)}}\rightarrow x_\text{m}^\dagger ]
        &= \frac{1}{N_{\text{m}}^{\text{(o)}}} 
        \\
        P_\text{sel}[X_{\text{h}}^\text{(o)}\rightarrow x_\text{h}^\dagger ]
        &= \frac{1}{N_{\text{h}}^\text{(o)}}
    \end{align}
    \label{eq:phase-point-selection}
\end{subequations}
Next, the Hamiltonians are swapped at the selected phase points. This is equivalent to swapping the phase points. This means
\begin{subequations}
    \begin{align}
    \text{main:}   \;& x_\text{m}^\dagger \rightarrow x_\text{h}^\dagger \\
    \text{helper:} \;& x_\text{h}^\dagger   \rightarrow x_\text{m}^\dagger
    \end{align}
    \label{eq:swap-forcefield}
\end{subequations}
\noindent Next, the new paths $X_{\text{m}}^{\text{(n)}}$ and $X_{\text{h}}^{\text{(n)}}$ are fully propagated from $x_\text{h}^\dagger$ and $x_\text{m}^\dagger$ by integrating the dynamics forward and backward in time, 
until the path reaches $\lambda_A$ or $\lambda_B$. In accordance with Eq.~\ref{eq:path-density}, the propagation probability that path $X_{\text{m}}^{\text{(n)}}$ is constructed from phase point $x_\text{h}^\dagger$ is
\begin{multline}
    P_\text{prop,m}
    [x_\text{h}^\dagger \rightarrow X_{\text{m}}^{\text{(n)}}] \\
    = \prod_{j=k}^{N_\text{m}^{\text{(n)}}-1} 
    p_\text{m}(x^{j} \rightarrow x^{j+1})
    \prod_{j=1}^{k} 
    p_\text{m}(\bar{x}^{j} \rightarrow \bar{x}^{j-1})
\end{multline}
where it is assumed that phase point $x_\text{h}^\dagger$ is the $k$-th phase point in the new path $X_{\text{m}}^{\text{(n)}}$ with length $N_\text{m}^\text{(n)}$.
Comparison with Eqs.~\ref{eq:path-density} and \ref{eq:phasepoint-density} shortens this propagation probability to
\begin{equation}
    P_\text{prop,m}
    [x_\text{h}^\dagger \rightarrow X_{\text{m}}^{\text{(n)}}]
    =  \frac{p_\text{m}[X_\text{m}^\text{(n)}]}{p_\text{m}(x_\text{h}^\dagger)} 
\end{equation}
with similar expressions for 
$P_\text{prop,h} [x_\text{m}^\dagger \rightarrow X_{\text{h}}^\text{(n)}]$.
The generation probability becomes
\begin{equation}
\begin{aligned}
    P_\text{gen} & (z^{\text{(o)}}\rightarrow z^{\text{(n)}})\\
    & =
    P_\text{sel}[X_{\text{m}}^{\text{(o)}}\rightarrow x_\text{m}^\dagger] \,
    P_\text{prop,m}[x_\text{h}^\dagger \rightarrow X_{\text{m}}^{\text{(n)}}] \\
    & \times
    P_\text{sel}[X_{\text{h}}^{\text{(o)}}\rightarrow x_\text{h}^\dagger] \,
    P_\text{prop,h}[x_\text{m}^\dagger \rightarrow X_{\text{h}}^{\text{(n)}}]
\end{aligned}
\label{eq:Pgen}
\end{equation}

The detailed balance condition is found from Eq.~\ref{eq:ratioPacc} by substituting
$P(z)$ of Eq.~\ref{eq:Pz} and $P_\text{gen}$ of Eq.~\ref{eq:Pgen}, and by making use of Eqs.~\ref{eq:phasepoint-density}-\ref{eq:path-density-lambda},
\begin{equation}
    \begin{aligned}
    \frac{P_\text{acc}(z^{\text{(o)}}\rightarrow z^{\text{(n)}})}
        {P_\text{acc}(z^{\text{(n)}}\rightarrow z^{\text{(o)}})}
    & = \frac{\mathds{1}_{i}[X_{\text{m}}^{\text{(n)}}]
          \mathds{1}_{i}[X_{\text{h}}^{\text{(n)}}]}
         {\mathds{1}_{i}[X_{\text{m}}^{\text{(o)}}]
          \mathds{1}_{i}[X_{\text{h}}^{\text{(o)}}]} \\
    & \times \frac{ N_{\text{m}}^{\text{(o)}}N_{\text{h}}^{\text{(o)}} }
          { N_{\text{m}}^{\text{(n)}}N_{\text{h}}^{\text{(n)}} }
      e^{-\beta \Delta\Delta \mathcal{H}}
    \end{aligned}
    \label{eq:acc-criterion-ratio}
\end{equation}
with
\begin{equation}
    \begin{aligned}
    \Delta\Delta \mathcal{H} & = \mathcal{H}_{\text{h}}(x_\text{m}^\dagger) - \mathcal{H}_{\text{h}}(x_\text{h}^\dagger)\\
    & - \mathcal{H}_{\text{m}}(x_\text{m}^\dagger) + \mathcal{H}_{\text{m}}(x_\text{h}^\dagger)
    \end{aligned}
\end{equation}
In conclusion, the following acceptance criterion for the Hamiltonian exchange in ensemble $[i+]$ satisfies Eq.~\ref{eq:acc-criterion-ratio} and thus ensures detailed balance,
\begin{multline}
    P_\text{acc}(z^{\text{(o)}}\rightarrow z^{\text{(n)}})
     = \mathds{1}_{i}[X_{\text{m}}^{\text{(n)}}]
    \mathds{1}_{i}[X_{\text{h}}^{\text{(n)}}] \\
     \times \min \left\{ 1, 
    \frac{ N_{\text{m}}^{\text{(o)}}N_{\text{h}}^{\text{(o)}} }
         { N_{\text{m}}^{\text{(n)}}N_{\text{h}}^{\text{(n)}} } 
         e^{-\beta  \Delta\Delta \mathcal{H}} \right\}
\label{eq:acc-criterion}
\end{multline}
Further, as the velocities of $x_\text{m}^\dagger$ and $x_\text{h}^\dagger$ are not changed during the engine-swap, the term $\Delta\Delta \mathcal{H}$ simplifies to the difference $\Delta\Delta U$ of potential energy shifts (Eq.~\ref{Eq:DeltaDeltaU}), giving the acceptance criterion in Eq.~\ref{eq:acc-criterion-method}.

\section*{Acknowledgements}

The authors would like to thank Titus S.\ van Erp and Lukas Baldauf for valuable discussions on the integration of our algorithm within the existing infinite-swap framework.
The computational resources (Stevin Supercomputer Infrastructure) and services used in this work were provided by the VSC (Flemish Supercomputer Center), funded by Ghent University, FWO and the Flemish Government – department EWI. 
SS and AG acknowledge funding of the FWO (projects G002520N, G094023N, and K222825N), BOF of Ghent University, and the European Union (ERC Consolidator grant, 101086145 PASTIME).

\section*{Data availability}

The data for all the simulations are publicly available on Zenodo at: 

https://doi.org/10.5281/zenodo.21025560

\bibliography{References}

@article{falkner2024enhanced,
  title={Enhanced sampling of configuration and path space in a generalized ensemble by shooting point exchange},
  author={Falkner, Sebastian and Coretti, Alessandro and Dellago, Christoph},
  journal={Physical Review Letters},
  volume={132},
  number={12},
  pages={128001},
  year={2024},
  publisher={APS}
}

@article{zhang2023enhanced,
  title={Enhanced path sampling using subtrajectory Monte Carlo moves},
  author={Zhang, Daniel T and Riccardi, Enrico and van Erp, Titus S},
  journal={The Journal of Chemical Physics},
  volume={158},
  number={2},
  pages={024113},
  year={2023},
  publisher={AIP Publishing}
}

@article{ghysels2021exact,
  title={Exact non-Markovian permeability from rare event simulations},
  author={Ghysels, An and Roet, Sander and Davoudi, Samaneh and van Erp, Titus S},
  journal={Physical Review Research},
  volume={3},
  number={3},
  pages={033068},
  year={2021},
  publisher={APS}
}

@article{souza2021martini,
  title={Martini 3: a general purpose force field for coarse-grained molecular dynamics},
  author={Souza, Paulo CT and Alessandri, Riccardo and Barnoud, Jonathan and Thallmair, Sebastian and Faustino, Ignacio and Gr{\"u}newald, Fabian and Patmanidis, Ilias and Abdizadeh, Haleh and Bruininks, Bart MH and Wassenaar, Tsjerk A and others},
  journal={Nature Methods},
  volume={18},
  number={4},
  pages={382--388},
  year={2021},
  publisher={Nature Publishing Group US New York}
}

@article{safaei2025exact,
  title={Exact Kinetics of Drug Permeation Using Transition Interface Sampling},
  author={Safaei, Sina and Baldauf, Lukas and van Erp, Titus S and Ghysels, An},
  journal={The Journal of Physical Chemistry B},
  volume={129},
  number={39},
  pages={10019--10034},
  year={2025},
  publisher={ACS Publications}
}

@article{jo2008charmm,
  title={CHARMM-GUI: a web-based graphical user interface for CHARMM},
  author={Jo, Sunhwan and Kim, Taehoon and Iyer, Vidyashankara G and Im, Wonpil},
  journal={Journal of Computational Chemistry},
  volume={29},
  number={11},
  pages={1859--1865},
  year={2008},
  publisher={Wiley Online Library}
}

@article{bussi2007canonical,
  title={Canonical sampling through velocity rescaling},
  author={Bussi, Giovanni and Donadio, Davide and Parrinello, Michele},
  journal={The Journal of Chemical Physics},
  volume={126},
  number={1},
  pages={014101},
  year={2007},
  publisher={AIP},
  doi={10.1063/1.2408420}
}

@article{bernetti2020pressure,
  title={Pressure control using stochastic cell rescaling},
  author={Bernetti, Mattia and Bussi, Giovanni},
  journal={The Journal of Chemical Physics},
  volume={153},
  number={11},
  pages={114107},
  year={2020},
  publisher={AIP Publishing}
}

@article{tong2019experimental,
  title={Experimental and molecular dynamics simulation study of the effects of lignin dimers on the gel-to-fluid phase transition in DPPC bilayers},
  author={Tong, Xinjie and Moradipour, Mahsa and Novak, Brian and Kamali, Poorya and Asare, Shardrack O and Knutson, Barbara L and Rankin, Stephen E and Lynn, Bert C and Moldovan, Dorel},
  journal={The Journal of Physical Chemistry B},
  volume={123},
  number={39},
  pages={8247--8260},
  year={2019},
  publisher={ACS Publications}
}

@article{roet2022exchanging,
  title={Exchanging replicas with unequal cost, infinitely and permanently},
  author={Roet, Sander and Zhang, Daniel T and van Erp, Titus S},
  journal={The Journal of Physical Chemistry A},
  volume={126},
  number={47},
  pages={8878--8886},
  year={2022},
  publisher={ACS Publications}
}

@article{van2026generalized,
  title={Generalized Path Reweighting and History-Dependent Free Energies},
  author={van Erp, Titus S and Zhang, Daniel T and Wils, Elias and Safaei, Sina and Ghysels, An},
  journal={The Journal of Chemical Physics},
  volume={164},
  number={21},
  pages={214107},
  year={2026},
  publisher={AIP Publishing}
}

@article{bolhuis2002transition,
  title={Transition path sampling: Throwing ropes over rough mountain passes, in the dark},
  author={Bolhuis, Peter G and Chandler, David and Dellago, Christoph and Geissler, Phillip L},
  journal={Annual Review of Physical Chemistry},
  volume={53},
  number={1},
  pages={291--318},
  year={2002},
  publisher={Annual Reviews 4139 El Camino Way, PO Box 10139, Palo Alto, CA 94303-0139, USA}
}

@article{dellago2008transition,
  title={Transition path sampling and other advanced simulation techniques for rare events},
  author={Dellago, Christoph and Bolhuis, Peter G},
  journal={Advanced Computer Simulation Approaches for Soft Matter Sciences III},
  volume={221},
  pages={167--233},
  year={2008},
  publisher={Springer}
}

@article{dellago1998efficient,
  title={Efficient transition path sampling: Application to Lennard-Jones cluster rearrangements},
  author={Dellago, Christoph and Bolhuis, Peter G and Chandler, David},
  journal={The Journal of Chemical Physics},
  volume={108},
  number={22},
  pages={9236--9245},
  year={1998},
  publisher={American Institute of Physics}
}

@article{dellago1998transition,
  title={Transition path sampling and the calculation of rate constants},
  author={Dellago, Christoph and Bolhuis, Peter G and Csajka, F{\'e}lix S and Chandler, David},
  journal={The Journal of Chemical Physics},
  volume={108},
  number={5},
  pages={1964--1977},
  year={1998},
  publisher={American Institute of Physics}
}

@article{van2003novel,
  title={A novel path sampling method for the calculation of rate constants},
  author={van Erp, Titus S and Moroni, Daniele and Bolhuis, Peter G},
  journal={The Journal of Chemical Physics},
  volume={118},
  number={17},
  pages={7762--7774},
  year={2003},
  publisher={American Institute of Physics}
}

@article{van2007reaction,
  title={Reaction rate calculation by parallel path swapping},
  author={van Erp, Titus S},
  journal={Physical Review Letters},
  volume={98},
  number={26},
  pages={268301},
  year={2007},
  publisher={APS}
}

@article{zhang2024highly,
  title={Highly parallelizable path sampling with minimal rejections using asynchronous replica exchange and infinite swaps},
  author={Zhang, Daniel T and Baldauf, Lukas and Roet, Sander and Lervik, Anders and van Erp, Titus S},
  journal={Proceedings of the National Academy of Sciences},
  volume={121},
  number={7},
  pages={e2318731121},
  year={2024},
  publisher={National Academy of Sciences}
}

@article{vervust2026estimating,
  title={Estimating full path lengths and kinetics from partial path transition interface sampling simulations},
  author={Vervust, Wouter and Wils, Elias and Safaei, Sina and Zhang, Daniel T and Ghysels, An},
  journal={Journal of Chemical Theory and Computation},
  volume={22},
  number={4},
  pages={1527--1538},
  year={2026},
  publisher={ACS Publications}
}

@article{vervust2025path,
  title={Path sampling challenges in large biomolecular systems: RETIS and REPPTIS for ABL-imatinib kinetics},
  author={Vervust, Wouter and Zhang, Daniel T and Riccardi, Enrico and van Erp, Titus S and Ghysels, An},
  journal={Biophysical Journal},
  volume={124},
  number={22},
  pages={3932--3947},
  year={2025},
  publisher={Elsevier}
}

@article{rogal2010reweighted,
  title={The reweighted path ensemble},
  author={Rogal, Jutta and Lechner, Wolfgang and Juraszek, Jarek and Ensing, Bernd and Bolhuis, Peter G},
  journal={The Journal of Chemical Physics},
  volume={133},
  number={17},
  pages={174109},
  year={2010},
  publisher={AIP Publishing}
}

@article{van2016analyzing,
  title={Analyzing complex reaction mechanisms using path sampling},
  author={van Erp, Titus S and Moqadam, Mahmoud and Riccardi, Enrico and Lervik, Anders},
  journal={Journal of Chemical Theory and Computation},
  volume={12},
  number={11},
  pages={5398--5410},
  year={2016},
  publisher={ACS Publications}
}

@article{riccardi2017fast,
  title={Fast decorrelating Monte Carlo moves for efficient path sampling},
  author={Riccardi, Enrico and Dahlen, Oda and van Erp, Titus S},
  journal={The Journal of Physical Chemistry Letters},
  volume={8},
  number={18},
  pages={4456--4460},
  year={2017},
  publisher={ACS Publications}
}

@article{bolhuis2008rare,
  title={Rare events via multiple reaction channels sampled by path replica exchange},
  author={Bolhuis, Peter G},
  journal={The Journal of Chemical Physics},
  volume={129},
  number={11},
  pages={114108},
  year={2008},
  publisher={AIP Publishing}
}

@article{falkner2023conditioning,
  title={Conditioning Boltzmann generators for rare event sampling},
  author={Falkner, Sebastian and Coretti, Alessandro and Romano, Salvatore and Geissler, Phillip L and Dellago, Christoph},
  journal={Machine Learning: Science and Technology},
  volume={4},
  number={3},
  pages={035050},
  year={2023},
  publisher={IOP Publishing}
}

@article{noe2019boltzmann,
  title={Boltzmann generators: Sampling equilibrium states of many-body systems with deep learning},
  author={No{\'e}, Frank and Olsson, Simon and K{\"o}hler, Jonas and Wu, Hao},
  journal={Science},
  volume={365},
  number={6457},
  pages={eaaw1147},
  year={2019},
  publisher={American Association for the Advancement of Science}
}

@article{hall2022practical,
  title={Practical guide to replica exchange transition interface sampling and forward flux sampling},
  author={Hall, Steven W and D{\'\i}az Leines, Grisell and Sarupria, Sapna and Rogal, Jutta},
  journal={The Journal of Chemical Physics},
  volume={156},
  number={20},
  pages={200901},
  year={2022},
  publisher={AIP Publishing}
}

@article{borrero2016avoiding,
  title={Avoiding traps in trajectory space: Metadynamics enhanced transition path sampling},
  author={Borrero, Ernesto E and Dellago, Christoph},
  journal={The European Physical Journal Special Topics},
  volume={225},
  number={8},
  pages={1609--1620},
  year={2016},
  publisher={Springer}
}

@article{bolhuis2021transition,
  title={Transition path sampling as Markov chain Monte Carlo of trajectories: Recent algorithms, software, applications, and future outlook},
  author={Bolhuis, Peter G and Swenson, David WH},
  journal={Advanced Theory and Simulations},
  volume={4},
  number={4},
  pages={2000237},
  year={2021},
  publisher={Wiley Online Library}
}

@article{van2005elaborating,
  title={Elaborating transition interface sampling methods},
  author={Van Erp, Titus S and Bolhuis, Peter G},
  journal={Journal of Computational Physics},
  volume={205},
  number={1},
  pages={157--181},
  year={2005},
  publisher={Elsevier}
}

@article{lervik2015gluing,
  title={Gluing potential energy surfaces with rare event simulations},
  author={Lervik, Anders and van Erp, Titus S},
  journal={Journal of Chemical Theory and Computation},
  volume={11},
  number={6},
  pages={2440--2450},
  year={2015},
  publisher={ACS Publications}
}

@article{hjorth2017atomic,
  title={The atomic simulation environment—a Python library for working with atoms},
  author={Hjorth Larsen, Ask and J{\o}rgen Mortensen, Jens and Blomqvist, Jakob and Castelli, Ivano E and Christensen, Rune and Du{\l}ak, Marcin and Friis, Jesper and Groves, Michael N and Hammer, Bj{\o}rk and Hargus, Cory and others},
  journal={Journal of Physics: Condensed Matter},
  volume={29},
  number={27},
  pages={273002},
  year={2017},
  publisher={IOP Publishing}
}

@article{vervust2024pyretis,
  title={PyRETIS 3: Conquering rare and slow events without boundaries},
  author={Vervust, Wouter and Zhang, Daniel T and Ghysels, An and Roet, Sander and van Erp, Titus S and Riccardi, Enrico},
  journal={Journal of Computational Chemistry},
  volume={45},
  number={15},
  pages={1224--1234},
  year={2024},
  publisher={Wiley Online Library}
}

@article{bolhuis2015practical,
  title={Practical and conceptual path sampling issues},
  author={Bolhuis, Peter G and Dellago, Christoph},
  journal={The European Physical Journal Special Topics},
  volume={224},
  number={12},
  pages={2409--2427},
  year={2015},
  publisher={Springer}
}

@article{chong2017path,
  title={Path-sampling strategies for simulating rare events in biomolecular systems},
  author={Chong, Lillian T and Saglam, Ali S and Zuckerman, Daniel M},
  journal={Current Opinion in Structural Biology},
  volume={43},
  pages={88--94},
  year={2017},
  publisher={Elsevier}
}

@article{van2023far,
  title={How far can we stretch the timescale with RETIS?},
  author={van Erp, Titus S},
  journal={Europhysics Letters},
  volume={143},
  number={3},
  pages={30001},
  year={2023},
  publisher={EDP Sciences, IOP Publishing and Societ{\`a} Italiana di Fisica}
}

@article{ferrenberg1989optimized,
  title={Optimized monte carlo data analysis},
  author={Ferrenberg, Alan M and Swendsen, Robert H},
  journal={Physical Review Letters},
  volume={63},
  number={12},
  pages={1195},
  year={1989},
  publisher={APS}
}

@article{kumar1992weighted,
  title={The weighted histogram analysis method for free-energy calculations on biomolecules. I. The method},
  author={Kumar, Shankar and Rosenberg, John M and Bouzida, Djamal and Swendsen, Robert H and Kollman, Peter A},
  journal={Journal of Computational Chemistry},
  volume={13},
  number={8},
  pages={1011--1021},
  year={1992},
  publisher={Wiley Online Library}
}

@article{roux1995calculation,
  title={The calculation of the potential of mean force using computer simulations},
  author={Roux, Beno{\^\i}t},
  journal={Computer Physics Communications},
  volume={91},
  number={1-3},
  pages={275--282},
  year={1995},
  publisher={Elsevier}
}

@article{zhang2024combining,
  title={Combining transition path sampling with data-driven collective variables through a reactivity-biased shooting algorithm},
  author={Zhang, Jintu and Zhang, Odin and Bonati, Luigi and Hou, TingJun},
  journal={Journal of Chemical Theory and Computation},
  volume={20},
  number={11},
  pages={4523--4532},
  year={2024},
  publisher={ACS Publications}
}

@article{rudzinski2019recent,
  title={Recent progress towards chemically-specific coarse-grained simulation models with consistent dynamical properties},
  author={Rudzinski, Joseph F},
  journal={Computation},
  volume={7},
  number={3},
  pages={42},
  year={2019},
  publisher={MDPI}
}

@article{fritz2011multiscale,
  title={Multiscale modeling of soft matter: scaling of dynamics},
  author={Fritz, Dominik and Koschke, Konstantin and Harmandaris, Vagelis A and van der Vegt, Nico FA and Kremer, Kurt},
  journal={Physical Chemistry Chemical Physics},
  volume={13},
  number={22},
  pages={10412--10420},
  year={2011},
  publisher={Royal Society of Chemistry}
}

@article{sadeghi2020large,
  title={Large-scale simulation of biomembranes incorporating realistic kinetics into coarse-grained models},
  author={Sadeghi, Mohsen and No{\'e}, Frank},
  journal={Nature Communications},
  volume={11},
  number={1},
  pages={2951},
  year={2020},
  publisher={Nature Publishing Group UK London}
}

@article{brasnett2025martiniglass,
  title={MartiniGlass: a tool for enabling visualization of coarse-grained martini topologies},
  author={Brasnett, Christopher and Marrink, Siewert J},
  journal={Journal of Chemical Information and Modeling},
  volume={65},
  number={7},
  pages={3137--3141},
  year={2025},
  publisher={ACS Publications}
}

@article{sugita2000multidimensional,
  title={Multidimensional replica-exchange method for free-energy calculations},
  author={Sugita, Yuji and Kitao, Akio and Okamoto, Yuko},
  journal={The Journal of Chemical Physics},
  volume={113},
  number={15},
  pages={6042--6051},
  year={2000},
  publisher={American Institute of Physics}
}

@article{unglert2025replica,
  title={Replica exchange nested sampling},
  author={Unglert, Nico and P{\'a}rtay, Livia Bart{\'o}k and Madsen, Georg Kent Hellerup},
  journal={Journal of Chemical Theory and Computation},
  volume={21},
  number={15},
  pages={7304--7319},
  year={2025},
  publisher={ACS Publications}
}

@article{fujisaki2010onsager,
  title={Onsager--Machlup action-based path sampling and its combination with replica exchange for diffusive and multiple pathways},
  author={Fujisaki, Hiroshi and Shiga, Motoyuki and Kidera, Akinori},
  journal={The Journal of Chemical Physics},
  volume={132},
  number={13},
  pages={134101},
  year={2010},
  publisher={AIP Publishing}
}

@article{sugita1999replica,
  title={Replica-exchange molecular dynamics method for protein folding},
  author={Sugita, Yuji and Okamoto, Yuko},
  journal={Chemical Physics Letters},
  volume={314},
  number={1-2},
  pages={141--151},
  year={1999},
  publisher={Elsevier}
}

@article{hukushima1996exchange,
  title={Exchange Monte Carlo method and application to spin glass simulations},
  author={Hukushima, Koji and Nemoto, Koji},
  journal={Journal of the Physical Society of Japan},
  volume={65},
  number={6},
  pages={1604--1608},
  year={1996},
  publisher={The Physical Society of Japan}
}

@article{bussi2014hamiltonian,
  title={Hamiltonian replica exchange in GROMACS: a flexible implementation},
  author={Bussi, Giovanni},
  journal={Molecular Physics},
  volume={112},
  number={3-4},
  pages={379--384},
  year={2014},
  publisher={Taylor \& Francis}
}

@article{okabe2001replica,
  title={Replica-exchange Monte Carlo method for the isobaric--isothermal ensemble},
  author={Okabe, Tsuneyasu and Kawata, Masaaki and Okamoto, Yuko and Mikami, Masuhiro},
  journal={Chemical Physics Letters},
  volume={335},
  number={5-6},
  pages={435--439},
  year={2001},
  publisher={Elsevier}
}

@article{klauda2010update,
  title={Update of the CHARMM all-atom additive force field for lipids: validation on six lipid types},
  author={Klauda, Jeffery B and Venable, Richard M and Freites, J Alfredo and O’Connor, Joseph W and Tobias, Douglas J and Mondragon-Ramirez, Carlos and Vorobyov, Igor and MacKerell Jr, Alexander D and Pastor, Richard W},
  journal={The Journal of Physical Chemistry B},
  volume={114},
  number={23},
  pages={7830--7843},
  year={2010},
  publisher={ACS Publications}
}

@article{chen2016multiple,
  title={Multiple time-step dual-Hamiltonian hybrid molecular dynamics--Monte Carlo canonical propagation algorithm},
  author={Chen, Yunjie and Kale, Seyit and Weare, Jonathan and Dinner, Aaron R and Roux, Beno{\^\i}t},
  journal={Journal of Chemical Theory and Computation},
  volume={12},
  number={4},
  pages={1449--1458},
  year={2016},
  publisher={ACS Publications}
}

@article{falkner2025revisiting,
  title={Revisiting shooting point Monte Carlo methods for transition path sampling},
  author={Falkner, Sebastian and Coretti, Alessandro and Peters, Baron and Bolhuis, Peter G and Dellago, Christoph},
  journal={The Journal of Chemical Physics},
  volume={163},
  number={3},
  pages={034105},
  year={2025},
  publisher={AIP Publishing}
}

@article{peters2006obtaining,
  title={Obtaining reaction coordinates by likelihood maximization},
  author={Peters, Baron and Trout, Bernhardt L},
  journal={The Journal of Chemical Physics},
  volume={125},
  number={5},
  pages={054108},
  year={2006},
  publisher={AIP Publishing}
}

@article{abrams2013enhanced,
  title={Enhanced sampling in molecular dynamics using metadynamics, replica-exchange, and temperature-acceleration},
  author={Abrams, Cameron and Bussi, Giovanni},
  journal={Entropy},
  volume={16},
  number={1},
  pages={163--199},
  year={2013},
  publisher={Molecular Diversity Preservation International (MDPI)}
}

@article{mullen2015easy,
  title={Easy transition path sampling methods: Flexible-length aimless shooting and permutation shooting},
  author={Mullen, Ryan Gotchy and Shea, Joan-Emma and Peters, Baron},
  journal={Journal of Chemical Theory and Computation},
  volume={11},
  number={6},
  pages={2421--2428},
  year={2015},
  publisher={ACS Publications}
}

@article{kang2026convergence,
  title={Convergence is not correctness: context-dependent performance of enhanced-sampling methods across biological complexity},
  author={Kang, Christopher and Chen, Cheng Giuseppe and Tang, Chenyu and Arredondo, Sergio Contreras and Zhou, Mengchen and Fu, Haohao and Yang, Lan and Gumbart, James C and Fakharzadeh, Ashkan and Moradi, Mahmoud and others},
  journal={Nature Communications},
  year={2026},
  publisher={Nature Publishing Group}
}

@article{asghar2024efficient,
  title={Efficient rare event sampling with unsupervised normalizing flows},
  author={Asghar, Solomon and Pei, Qing-Xiang and Volpe, Giorgio and Ni, Ran},
  journal={Nature Machine Intelligence},
  volume={6},
  number={11},
  pages={1370--1381},
  year={2024},
  publisher={Nature Publishing Group UK London}
}

@article{wang2022efficient,
  title={Efficient sampling of high-dimensional free energy landscapes using adaptive reinforced dynamics},
  author={Wang, Dongdong and Wang, Yanze and Chang, Junhan and Zhang, Linfeng and Wang, Han and E, Weinan},
  journal={Nature Computational Science},
  volume={2},
  number={1},
  pages={20--29},
  year={2022},
  publisher={Nature Publishing Group US New York}
}

@article{robo2023fast,
  title={Fast free energy estimates from $\lambda$-dynamics with bias-updated Gibbs sampling},
  author={Robo, Michael T and Hayes, Ryan L and Ding, Xinqiang and Pulawski, Brian and Vilseck, Jonah Z},
  journal={Nature Communications},
  volume={14},
  number={1},
  pages={8515},
  year={2023},
  publisher={Nature Publishing Group UK London}
}

@article{li2025enhanced,
  title={Enhanced sampling of protein conformational changes via true reaction coordinates from energy relaxation},
  author={Li, Huiyu and Ma, Ao},
  journal={Nature Communications},
  volume={16},
  number={1},
  pages={786},
  year={2025},
  publisher={Nature Publishing Group UK London}
}

@article{post2025ai,
  title={AI-guided transition path sampling of lipid flip-flop and membrane nanoporation},
  author={Post, Matthias and Hummer, Gerhard},
  journal={Nature Communications},
  pages={224},
  number={1},
  year={2025},
  publisher={Nature Publishing Group UK London}
}

@article{trizio2025everything,
  title={Everything everywhere all at once: a probability-based enhanced sampling approach to rare events},
  author={Trizio, Enrico and Kang, Peilin and Parrinello, Michele},
  journal={Nature Computational Science},
  volume={5},
  number={7},
  pages={582--591},
  year={2025},
  publisher={Nature Publishing Group US New York}
}

@article{plattner2017complete,
  title={Complete protein--protein association kinetics in atomic detail revealed by molecular dynamics simulations and Markov modelling},
  author={Plattner, Nuria and Doerr, Stefan and De Fabritiis, Gianni and No{\'e}, Frank},
  journal={Nature Chemistry},
  volume={9},
  number={10},
  pages={1005--1011},
  year={2017},
  publisher={Nature Publishing Group UK London}
}

@article{plattner2015protein,
  title={Protein conformational plasticity and complex ligand-binding kinetics explored by atomistic simulations and Markov models},
  author={Plattner, Nuria and No{\'e}, Frank},
  journal={Nature Communications},
  volume={6},
  number={1},
  pages={7653},
  year={2015},
  publisher={Nature Publishing Group UK London}
}

@article{frallicciardi2022determining,
  title={Determining small-molecule permeation through lipid membranes},
  author={Frallicciardi, Jacopo and Gabba, Matteo and Poolman, Bert},
  journal={Nature Protocols},
  volume={17},
  number={11},
  pages={2620--2646},
  year={2022},
  publisher={Nature Publishing Group UK London}
}

@article{haloi2025adaptive,
  title={Adaptive sampling--based structural prediction reveals opening of a GABAA receptor through the $\alpha$$\beta$ interface},
  author={Haloi, Nandan and Eriksson Lidbrink, Samuel and Howard, Rebecca J and Lindahl, Erik},
  journal={Science Advances},
  volume={11},
  number={2},
  pages={eadq3788},
  year={2025},
  publisher={American Association for the Advancement of Science}
}

@article{kang2024computing,
  title={Computing the committor with the committor to study the transition state ensemble},
  author={Kang, Peilin and Trizio, Enrico and Parrinello, Michele},
  journal={Nature Computational Science},
  volume={4},
  number={6},
  pages={451--460},
  year={2024},
  publisher={Nature Publishing Group US New York}
}

@article{casert2024learning,
  title={Learning stochastic dynamics and predicting emergent behavior using transformers},
  author={Casert, Corneel and Tamblyn, Isaac and Whitelam, Stephen},
  journal={Nature Communications},
  volume={15},
  number={1},
  pages={1875},
  year={2024},
  publisher={Nature Publishing Group UK London}
}

@article{elber2020milestoning,
  title={Milestoning: An efficient approach for atomically detailed simulations of kinetics in biophysics},
  author={Elber, Ron},
  journal={Annual Review of Biophysics},
  volume={49},
  number={1},
  pages={69--85},
  year={2020},
  publisher={Annual Reviews}
}

@article{tsai2022path,
  title={Path sampling of recurrent neural networks by incorporating known physics},
  author={Tsai, Sun-Ting and Fields, Eric and Xu, Yijia and Kuo, En-Jui and Tiwary, Pratyush},
  journal={Nature Communications},
  volume={13},
  number={1},
  pages={7231},
  year={2022},
  publisher={Nature Publishing Group UK London}
}

@article{wang2023predicting,
  title={Predicting biomolecular binding kinetics: A review},
  author={Wang, Jinan and Do, Hung N and Koirala, Kushal and Miao, Yinglong},
  journal={Journal of Chemical Theory and Computation},
  volume={19},
  number={8},
  pages={2135--2148},
  year={2023},
  publisher={ACS Publications}
}

@article{horvath2025stim1,
  title={STIM1 transmembrane helix dimerization captured by AI-guided transition path sampling},
  author={Horvath, Ferdinand and Jung, Hendrik and Grabmayr, Herwig and Fahrner, Marc and Romanin, Christoph and Hummer, Gerhard},
  journal={Proceedings of the National Academy of Sciences},
  volume={122},
  number={35},
  pages={e2506516122},
  year={2025},
  publisher={National Academy of Sciences}
}

@article{martinotti2020molecular,
  title={Molecular dynamics simulation of small molecules interacting with biological membranes},
  author={Martinotti, Carlo and Ruiz-Perez, Lanie and Deplazes, Evelyne and Mancera, Ricardo L},
  journal={ChemPhysChem},
  volume={21},
  number={14},
  pages={1486--1514},
  year={2020},
  publisher={Wiley Online Library}
}

@article{henin2022enhanced,
  title={Enhanced Sampling Methods for Molecular Dynamics Simulations},
  author={Hénin, Jérôme and Lelièvre, Tony and Shirts, Michael R. and Valsson, Omar and Delemotte, Lucie},
  journal={Living Journal of Computational Molecular Science},
  volume={4},
  number={1},
  pages={1583},
  year={2022},
}

@article{zhang2024alchemical,
  title={Alchemical enhanced sampling with optimized phase space overlap},
  author={Zhang, Shi and Giese, Timothy J and Lee, Tai-Sung and York, Darrin M},
  journal={Journal of Chemical Theory and Computation},
  volume={20},
  number={9},
  pages={3935--3953},
  year={2024},
  publisher={ACS Publications}
}

@misc{GROMACS,
  title        = {{GROMACS} 2024.4 Manual},
  author       = {{GROMACS development team}},
  year         = {2024},
  howpublished = {Software documentation},
  note         = {Version 2024.4},
  doi          = {10.5281/zenodo.14016590},
  url          = {https://doi.org/10.5281/zenodo.14016590}
}

@misc{infretissoftware,
  author       = {{infRETIS Development Team}},
  howpublished = {\url{https://github.com/infretis/}},
  year         = {2026},
  note         = {GitHub repository}
}

@misc{hretis_git,
  author       = {{HRETIS Development Team}},
  howpublished = {\url{https://github.com/SinaSina92/hretis}},
  year         = {2026},
  note         = {GitHub repository}
}

\end{document}